\documentclass[graphics, twocolumn, usenatbib]{mn2e}
\usepackage{times}
\usepackage{natbib} 
\usepackage{epsfig} 
\usepackage{graphicx} 
\usepackage{color}
\usepackage{aas_macros} 
\usepackage{amssymb}
\usepackage{amsmath}
\usepackage[title]{appendix}
\usepackage{hyperref}	% Hyperlinks
\hypersetup{colorlinks=true,linkcolor=blue,citecolor=blue,filecolor=blue,urlcolor=blue}
\usepackage[caption=false]{subfig}

\newcommand{\enzo}{\texttt{Enzo~}}
\newcommand{\enzoc}{\texttt{Enzo}}

\newcommand{\kms} {km $\rm{s^{-1}}$}
\newcommand{\mpch} {\rm $h^{-1}$ Mpc\,\,} 
\newcommand{\kpch} {\rm $h^{-1}$ kpc\,\,} 
\newcommand{\msolar} {$\rm{M_{\odot}}~$}
\newcommand{\msolarc} {$\rm{M_{\odot}}$}

\newcommand{\molH} {$\rm{H_2}$~}
\newcommand{\molHc} {$\rm{H_2}$}

\newcommand{\smartstar} {\texttt{SmartStar~}}
\newcommand{\smartstars} {\texttt{SmartStars~}}
\newcommand{\smartstarsc} {\texttt{SmartStars}}
\newcommand{\smartstarc} {\texttt{SmartStar}}

\begin{document}
\title[]{Rise of the First Super-Massive Stars}

\author[J.A. Regan \& T.P. Downes] 
{John A. Regan\thanks{E-mail:john.regan@dcu.ie, Marie Sk\l odowska-Curie Fellow} \& Turlough P. Downes \\ \\
Centre for Astrophysics \& Relativity, School of Mathematical Sciences, Dublin City University, Glasnevin, Ireland\\
\\}

%Start things off
%\date{Accepted XXX. Received YYY; in original form ZZZ}
\pubyear{2016}
\label{firstpage}
\pagerange{\pageref{firstpage}--\pageref{lastpage}}

%Make the Title
\maketitle

 %%%%%%%%%%%%%%%%%%%%%%%%%%%%%%%%%%%%%%%%%%%%%%%%%%%
%Abstract time
\begin{abstract} 
  We use high resolution adaptive mesh refinement simulations to model the formation
  of massive metal-free stars in the early Universe. By applying Lyman-Werner (LW) backgrounds of
  100 J$_{21}$ and 1000 J$_{21}$ respectively we construct environments conducive to
  the formation of massive stars. We find that only in the case of the higher LW backgrounds
  are super-critical accretion rates realised that are necessary for super-massive star (SMS)
  formation.  Mild fragmentation is observed for both backgrounds. Violent dynamical interactions
  between the stars that form in the more massive halo formed (1000 J$_{21}$ background) results in the
  eventual expulsion of the two most massive stars from the halo.  In the smaller mass halo (100
  J$_{21}$ background) mergers of stars occur before any multibody interactions and a single
  massive Pop III star is left at the centre of the halo at the end of our simulation. 
  Feedback from the very massive Pop III stars is not effective in generating a large HII region
  with ionising photons absorbed within a few thousand AU of the star.  In all cases a massive black
  hole seed is the expected final fate of the most massive objects. The
  seed of the massive Pop III star which remained at the centre of the less massive halo
  experiences steady accretion rates of almost  $10^{-2}$ \msolarc/yr and if these rates
  continue could potentially experience super-Eddington accretion rates in the immediate aftermath
  of collapsing into a black hole. 
\end{abstract}

%%%%%%%%%%%%%%%%%%%%%%%%%%%%%%%%%%%%%%%%%%%%%%%%%%%
%keywords time
%%%%%%%%%%%%%%%%%%%%%%%%%%%%%%%%%%%%%%%%%%%%%%%%%%%
%Introduction time

%%%%%%%%%%%%%%%%%%%%%%%%%%%%%%%%%%%%%%%%%%%%%%%%%%%%%%%%%%%%%%%%%%%%%%%%%%%%

\section{Introduction} \label{Sec:Introduction}
%%%%%%%%%%%%%%%%%%%%%%%%%%%%%%%Table 2%%%%%%%%%%%%%%%%%%%%%%%%%%%%%%%%%%%%%%%%%%
\begin{table*} 
\centering
\caption{Simulation Parameters}
\begin{tabular}{ | l | c | c | c | c | c | c |}
\hline 
\textbf{\em {Sim Name$^{a}$}} &
\textbf{\em {J$_{21}^b$}} & \textbf{\em Maximum Resolution$^{c}$ (pc)} &
\textbf{\em{Collapse Redshift$^{d}$}}
& \textbf{\em{M$_{\rm{core}}^f$ (\msolarc)}}
& \textbf{\em{M$_{\rm{halo}}^g$ (\msolarc)}}\\
\hline
Ref18\_1J       & 1     & 0.001 & 27.1 & 7740 & $9.56 \times 10^6$ \\
Ref10\_100J     & 100.0 & 0.44 & z = 24.7  & 16866 & $1.38\times 10^7$ \\
Ref14\_100J21   & 100.0 & 0.02 & z = 24.7  & 16462 & $1.38 \times 10^7$  \\
Ref18\_100J21   & 100.0 & 0.001 & z = 24.7  & 21194 & $1.39 \times 10^7$  \\
Ref20\_100J21\_OT   & 100.0 & 0.00025 & z = 24.7  & 21187 & $1.39 \times 10^7$  \\
%1000J\_Ref10     & 1000.0 & 0.44 & z = 24.6  & 2186 & $1.07 \times 10^6$ \\
Ref14\_1000J21   & 1000.0 & 0.02 & z = 23.7  & 58948 & $2.67 \times 10^7$  \\
Ref16\_1000J21   & 1000.0 & 0.004 & z = 23.7  & 59018 & $2.67 \times 10^7$  \\
Ref18\_1000J21\_OT   & 1000.0 & 0.001 & z = 23.7  & 59921 & $2.59 \times 10^7$  \\

\hline

\end{tabular}
\parbox[t]{0.9\textwidth}{\textit{Notes:} The details of each of the
  realisations used in this study. (a) The simulation name, (b) The LW intensity in units of J$_{21}$,
  (c) the maximum physical resolution, (d) the collapse redshift (i.e. the redshift at which the
  first smartstar forms),
  (f) gas mass within the central 1 pc just before the first \smartstar forms
  (g) is the halo virial mass  (DM \& gas) in solar masses.  
}

\label{Table:Sims}
\end{table*}

%%%%%%%%%%%%%%%%%%%%%%%%%%%%%%%%%%%%%%%%%%%%%%%%%%%%%%%%%%%%%%%%%%%%%%%%%%%%%%%%%%%%%%%%%%%%%
Very high redshift quasars powered by super massive black holes (SMBHs) have presented a
problem when trying to reconcile the very large masses of these objects with possible
progenitors. The most recent discovery being made by \cite{Banados_2018} of a quasar at z = 7.54
with an estimated mass of M = $8 \times 10^8$ \msolarc. This discovery adds to the previous
record holder of z = 7.085 \citep{Mortlock_2011}. At the present time there have been more than
one hundred quasars discovered at redshifts greater than 6 \citep{Wang_2016, Wang_2017}.
The presence of such massive objects at early times in the Universe presents a fundamental
problem in terms of the progenitors of these objects.  \\
\indent Stellar mass black holes formed from the
remnants of the first stars (Pop III stars) must accrete at the Eddington limit for their
entire history to reach the billion mass threshold by a redshift of 7 if they are to be the seeds
of the first quasars. This scenario appears exceedingly difficult as the first stellar
mass black holes are born ``starving'' \citep{Whalen_2004, Alvarez_2009, Johnson_2011}
in mini haloes which have been disrupted by both the ionising radiation from Pop III stars and the
subsequent supernova explosions \citep{Milosavljevic_2009, Jeon_2014}. Furthermore, the remnants
of Pop III stars continue to accrete very inefficiently even after migrating into larger haloes
via mergers (Smith et al. in prep).
As a direct result, investigation of supermassive star (SMS) formation as a viable alternative has
been undertaken and appears attractive \citep{Haiman_2006, Begelman_2006, Wise_2008a, 
  Regan_2009b, Regan_2009, Volonteri_2010a, Agarwal_2012, Agarwal_2013, Agarwal_2014b,
  Latif_2015b, Latif_2016a, Regan_2017}. \\
\indent Recent work on the formation of SMSs \citep{Woods_2017, Haemmerle_2017, Haemmerle_2017b,
  Hosokawa_2015, Hosokawa_2012, Hosokawa_2013} has come to the broad consensus that when proto-stars
are subject to accretion rates greater than approximately 0.04 \msolarc/yr the envelope surrounding
the proto-star becomes bloated leading to effective surface temperatures of approximately 5000 K.
SMSs are therefore expected to resemble red giants albeit with significantly more luminosity.
Using 2-D hydrodynamical simulations \cite{Sakurai_2016} show that even in the case when the
accretion rate onto the proto-star drops below the critical accretion rate the resulting UV feedback
is too weak to halt accretion as long as the periods of quiescence are less than approximately 1000
years. SMSs are then expected to continue accreting until they directly collapse into a black hole
with initial masses of approximately $\rm{M_{BH}} \sim 10^5$ \msolarc.\\
\indent Creating the conditions for generating these extreme accretion rates has therefore been
a topic of great interest to astrophysicists over the last decade or more. In the course of the
``normal'' collapse of structure, metal-free stars made entirely of Hydrogen and Helium form
with typical accretion rates of $\rm{M_{acc}} \lesssim 10^{-3}$ \msolarc/yr. Accretion rates of this
magnitude do not lead to SMS formation but rather the formation of Population III stars which are
extremely luminous in the UV. The feedback from the creation of the Pop III star effectively halts
accretion and is expected to set a characteristic mass of approximately 40 solar masses for
Pop III stars \citep{Hosokawa_2011, Hirano_2014, Hosokawa_2015}. However, as noted above black hole progenitors beginning with
initial masses of this magnitude must accrete at the Eddington rate for approximately a billion years
in order to achieve the masses required of the first quasars. \\
\indent As a result of the early bottlenecks likely suffered by the first stars in terms of their
growth prospects many authors have considered the idea that SMSs may be a more natural
progenitor for the first quasars. Their significantly increased initial masses alleviates some of the
difficulties in achieving final masses of a billion solar masses within a gigayear timeframe. 
The very high accretion rates required to form a SMS requires larger halos than those in
which normal Pop III star formation occurs. If star formation can be suppressed until
a dark matter halo reaches the atomic cooling threshold then the accretion rates onto a central
object can reach rates of nearly 1 \msolarc/yr - well in excess of the critical rate required.
The conditions required to form pristine atomic cooling haloes has been
investigated by several authors over the past decade or more. As Pop III stars form they emit copious
amounts of Lyman-Werner (LW) radiation. LW photons can dissociate \molH thus removing a crucial
coolant that allows Pop III star formation to proceed. If the LW flux impinging onto a growing halo
is large enough then Pop III star formation can be completely suppressed until the atomic cooling
limit is reached at which point cooling by atomic hydrogen drives the collapse. This mechanism
by which external (LW) radiation from one (or more) star forming haloes sterilises
a neighbouring
halo also requires that the target halo remains metal-free. As a result the term ``synchronised''
haloes has been coined \citep{Dijkstra_2008, Agarwal_2014b, Dijkstra2014a,
  Visbal_2014b, Regan_2017}. \\
\indent Another promising avenue to achieving the required accretion rates in pristine haloes is in
haloes that are exposed to large streaming velocities\citep{Tseliakhovich_2010}.
In this case the large offset in velocities
between the dark matter particles and the baryons suppresses the ability of the halo to virialise
until it's mass exceeds the atomic cooling threshold. Streaming velocities therefore also suppress
Pop III star formation in minihaloes. In this case it has been shown that as the halo collapses it
forms \molH but the accretion rates are so high that SMS formation may nonetheless
result \citep{Hirano_2017}. \cite{Schauer_2017} showed that streaming velocities may also provide
ideal sites for synchronization of haloes to take place with the first proto-galaxy to collapse
effectively sterilising the second galaxy promoting the formation of a SMS. \\
\indent In this study we explore the formation of a massive proto-star when exposed to large LW
backgrounds. The LW background allows us to suppress the formation of Pop III stars in minihaloes
in our volume with collapse only able to proceed once atomic cooling becomes important. We use
star particles to follow the evolution of proto-star formation. We explicitly track the accretion onto
the star particles and investigate if accretion rates above the critical threshold can be sustained
over the 1 Myr expected lifetime of the proto-star. At the end of its life the massive star directly
collapses into a black hole. In a companion paper we investigate the initial growth of the embryonic
black hole under different accretion and feedback regimes. \\
\indent The paper is laid out as follows: in \S \ref{Sec:Model} we describe the 
model setup and the numerical approach used as well as describing our star particle formulation;
in \S \ref{Sec:Results} we describe the results of our numerical simulations;
in \S \ref{Sec:Discussion} we discuss the importance of the results and present our conclusions.  \\
Throughout this paper we  assume a standard $\Lambda$CDM cosmology with the following parameters 
\cite[based on the latest Planck data]{Planck_2014}, $\Omega_{\Lambda,0}$  = 0.6817, 
$\Omega_{\rm m,0}$ = 0.3183, $\Omega_{\rm b,0}$ = 0.0463, $\sigma_8$ = 0.8347 and $h$ = 0.6704. 
We further assume a spectral index for the primordial density fluctuations of $n=0.9616$.

\section{Numerical Framework} \label{Sec:Model}
\noindent In this study we have used the publicly available adaptive mesh refinement code
\enzoc\footnote{http://enzo-project.org/} to study the fragmentation properties of gas within
haloes irradiated by a background LW field. Into \enzo we have added a new star particle type
which we have dubbed \smartstarc. We now describe both components.
\subsection{Enzo}
\enzoc\footnote{Changeset:fedb30ff370b} \citep{Enzo_2014} is an adaptive mesh refinement code
ideally suited for simulations of the high redshift universe. Gravity in \enzo is solved using
a fast Fourier technique \citep{Hockney_1988} which solves the Poisson equation on the root grid
at each timestep. On subgrids, the boundary
conditions are interpolated to the subgrids and the Poisson equation is then solved at each timestep.
Dark matter is represented using particles, each particle is stored on the highest refinement grid
available to it and thus the particle has the same timestep as the gas on that grid. The
particle densities are interpolated onto the grid and solved at the same time as the gas potential.
\enzo contains several hydrodynamics schemes to solve the Euler equation. We use the piecewise
parabolic method which was originally developed by \cite{Berger_1984} and adapted to cosmological
flows by \cite{Bryan_1995}. The PPM solver is an explicit, higher order accurate version of
Godunov's method for ideal gas dynamics with a spatially third accurate piecewise parabolic
monotonic interpolation scheme employed. A nonlinear Riemann solver is used for shock capturing. The
method is formally second order accurate in space and time and explicitly conserves mass, linear
momentum and energy making the scheme extremely useful for following the collapse of dense
structures. \\
\indent Chemistry is an important component in following the collapse of (ideal) gas. We use the
\texttt{Grackle}\footnote{https://grackle.readthedocs.org/}$^,$\footnote{Changeset:482876c71f73}
\citep{Grackle} library to follow the evolution of ten individual species:
${\rm H}, {\rm H}^+, {\rm He}, {\rm He}^+,  {\rm He}^{++}, {\rm e}^-,$ 
$\rm{H_2}, \rm{H_2^+}\, \rm{H^-} \rm{and}\ \rm{HeH^+}$. We adopt here the 26 reaction network
determined by \cite{Glover_2015a} as the most appropriate network for solving the chemical
equations required by gas of primordial composition with no metal pollution and exposed to an external
radiation source. The network includes the most 
up-to-date rates as described in \cite{GloverJappsen_2007},  \cite{GloverAbel_2008},
\cite{GloverSavin_2009}, \cite{Coppola_2011}, \cite{Coppola_2012},  \cite{Glover_2015a},
\cite{Glover_2015b},  \cite{Latif_2015}. The cooling mechanisms
included in the model are collisional excitation cooling, collisional ionisation cooling,
recombination cooling, bremsstrahlung and Compton cooling off the CMB.\\
\subsection{Simulation Setup}
\noindent The simulation volumes considered here are designed to explore the collapse
of a single cosmological halo with no metals. External LW backgrounds are used to suppress
\molH formation leading to the formation of pristine atomic cooling haloes. We now describe
the details of the setup. 
All simulations are run within a cosmological box of 2 \mpch (comoving), 
  the root grid size is $256^3$ and we employ three levels of nested grids. The grid nesting and
  initial conditions were created using MUSIC \citep{Hahn_2011}. Within the most refined region
  (i.e. level 3) the dark matter particle mass is $\sim$ 103 \msolarc. In order to increase further
  the dark matter resolution of our simulations we split the dark matter particles according to the
  prescription of \cite{Kitsionas_2002} and as described in \cite{Regan_2015}. We split particles
  centered on the position of the final collapse as found from lower resolution simulations within a
  region with a comoving side length of 43.75 h$^{-1}$ kpc. Each particle is split into 13 daughter
  particles resulting in a final high resolution region with a dark matter particle mass of
  $\sim$ 8 \msolarc. The particle splitting is done at a redshift of 40 well before the collapse of
  the target halo. Convergence testing to study the impact of lower dark matter particle masses was
  discussed in \cite{Regan_2015}. \\
  \indent The baryon resolution is set by the size of the grid cells. In the highest resolution region 
  this corresponds to approximately 0.48  \kpch comoving (before adaptive refinement). We vary
  the maximum refinement level (see Table \ref{Table:Sims}) to explore the impact of resolution on
  our results. Refinement is triggered in \enzo  when certain, user defined, thresholds are exceeded. The
  refinement criteria used in this work were based on three physical measurements: (1) The dark
  matter particle over-density, (2) The baryon over-density and (3) the Jeans length. The first two
  criteria introduce additional meshes when the over-density of a grid cell with respect to the
  mean gas or dark matter density exceeds 8.0. Furthermore,
  we set the \emph{MinimumMassForRefinementExponent} parameter
  to $-0.1$ making the refinement more aggressive for the baryon and dark matter over-density
    and hence making the behaviour of the adaptive mesh ``super-Lagrangian'' in nature (see \cite{Enzo_2014} for further details).
  This technique also reduces the threshold for
  refinement as higher densities are reached. For the final criteria we set the number of cells
  per Jeans length to be 32 in these runs. \\
  \indent We use between 10 and 20 levels of refinement in our simulations. A refinement level of
  10 corresponds to a comoving spatial resolution $\Delta x_c \sim 10$ pc and a physical resolution of
  $\Delta x_p \sim 0.4$ pc at z = 25. A refinement level of 20 on the other hand reduces our
  spatial resolution to $\Delta x_c \sim 0.01$ pc (comoving) and a physical resolution of
  $\Delta x_p \sim 0.0004$ pc ($\sim 90$ AU) at z = 25. At this resolution scale we are able to marginally
  resolve the formation of individual SMSs. In order to suppress Pop III star formation and allow the
  simulation to form pristine atomic cooling haloes we impose an artificial Lyman-Werner background.
  We set the effective temperature of the background radiation field to
  T$\rm{_{eff}} = 30000$ K. This background temperature suitably models the spectrum of a
  population of young stars \citep{WolcottGreen_2012, Sugimura_2014, Latif_2015}.
  The effective temperature of the background is important as the radiation temperature
  determines the dominant photo-dissociation reaction set in the irradiated halo. This in turn
  leads to a value of J$_{crit}$ - the flux above which complete isothermal collapse of the
  irradiated halo is observed due to the complete suppression of \molHc. The actual value of
  J$_{crit}$ depends on the nature of the source spectrum \citep{Shang_2010, Sugimura_2014,
    Agarwal_2015a}. We do not investigate the exact value of J$_{crit}$ here in SMS formation,
  rather we use the LW background as a mechanism to form atomic cooling haloes in which
  to probe SMS formation and evolution.\\
  %% \indent \cite{Agarwal_2015b} proposed that the J$_{crit}$ needed from a given stellar
  %% population modelled using realistic stellar spectra can vary widely over 2-3 orders of
  %% magnitude. They argue that in an external pristine atomic cooling halo, DCBH formation is
  %% better parameterised by using a critical curve in the H$_2$ and H$^-$ photo-destruction rate
  %% parameter space (further confirmed by \citep{WolcottGreen_2017}). While our choice of
  %% T$\rm{_{eff}} = 30000$ K falls well within the
  %% range advocated by these studies, including a realistic source spectrum derived form
  %% population synthesis models is beyond the scope of the current work and remains to be
  %% explored in a future study. In should also be noted that \cite{Sugimura_2014} found that the
  %% J$_{crit}$ is only very weakly dependent on the nature of the source spectrum in tension with
  %% the results of both \cite{Agarwal_2015b} and \citep{WolcottGreen_2017}, however they explored a
  %% somewhat smaller parameter space.
  %%%%%%%%%%%%%%%%%FIGURE 1%%%%%%%%%%%%%%%%%%%%%%%%%%%%%%%%%%%%%%%%%%%%%%%
\begin{figure*}
  \centering 
  \begin{minipage}{175mm}      \begin{center}
      \centerline{
        \includegraphics[width=9cm]{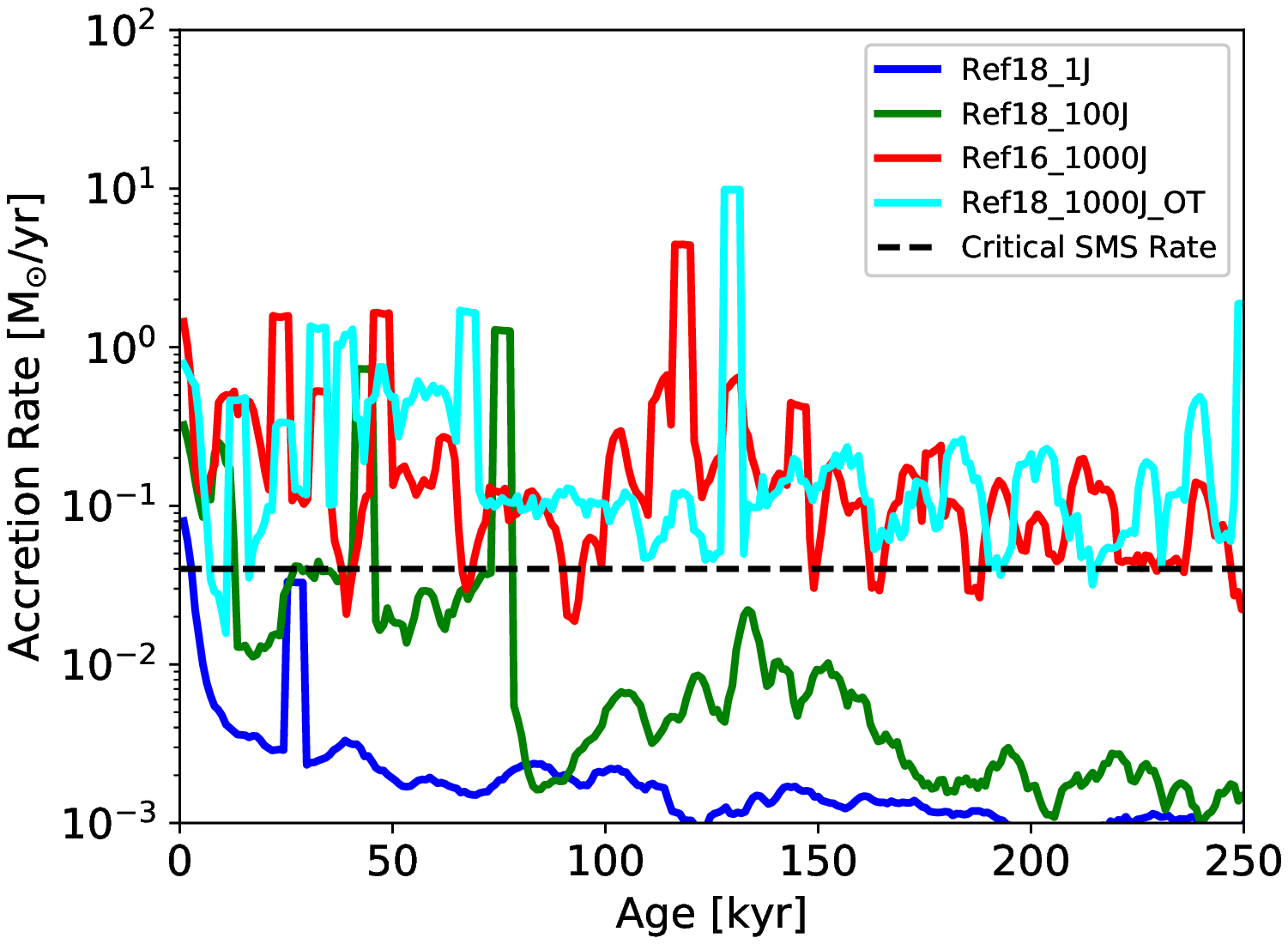}
        \includegraphics[width=9cm]{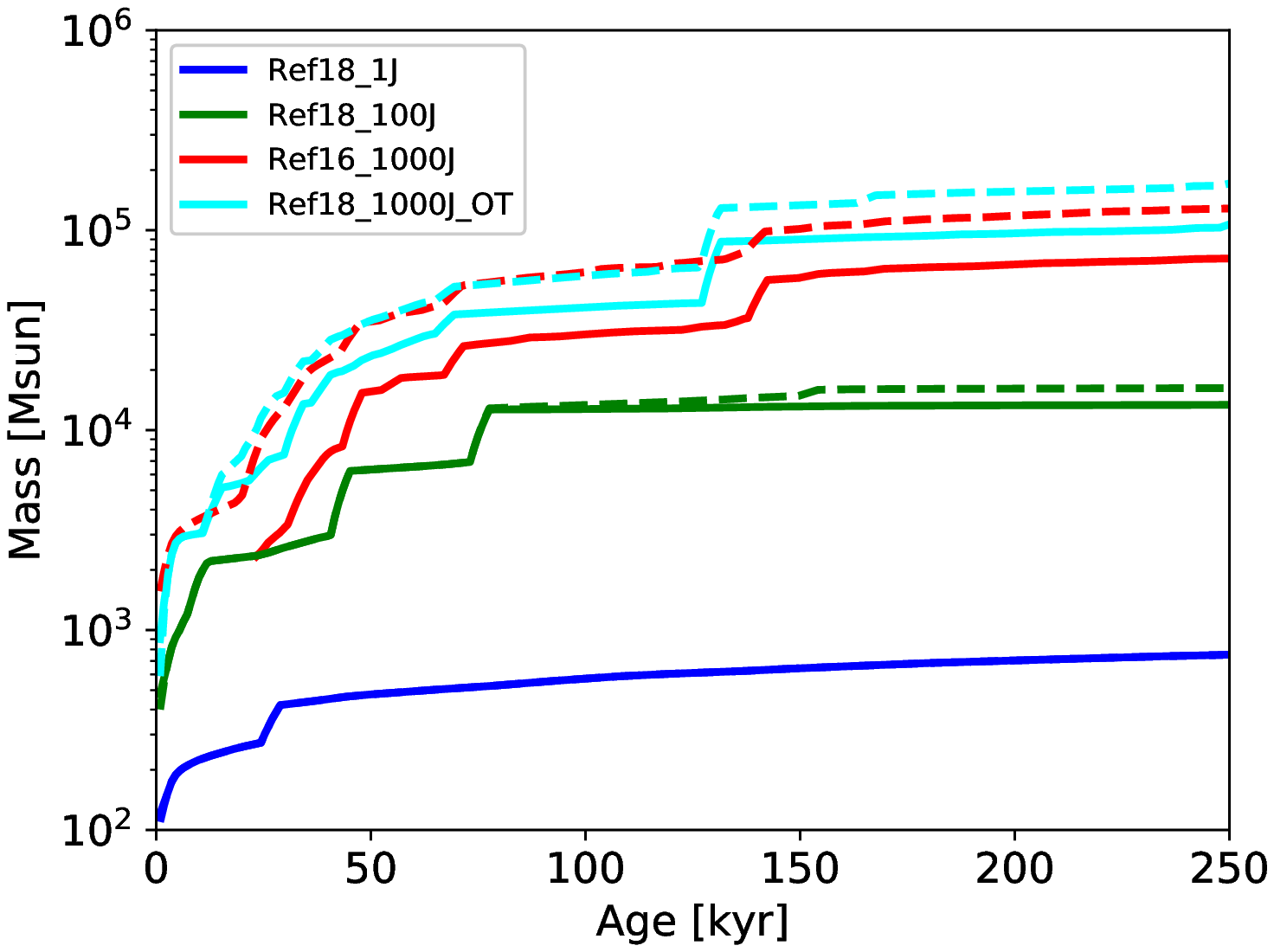}}
        \caption[]
        {\label{MassAccretionRate}
          The mass accretion rate onto the largest \smartstar particle in the simulation with
          a maximum refinement level of 18 (corresponding to a spatial resolution of
          $\Delta x \sim 0.001$ pc). For the Ref18\_1000J\_OT simulation the radiative transfer
          is made using the optically thin approximation and so we also include the Ref16\_1000J
          simulation (with full radiative transfer) for direct comparison. The accretion rates,
          on average, increase with the intensity of the background as expected. The convergence
          between the Ref16\_1000J and Ref18\_1000J\_OT results should also be noted. 
          In the right hand panel we show the mass of the largest particle in each case and the total
          mass of all \smartstar particles (dashed line). The 1 J$_{21}$ simulation shows the lowest
          stellar masses with the masses of the most massive stars increasing with background
          intensity. Again note the excellent convergence between the Ref16\_1000J and
          Ref18\_1000J\_OT results with differences within a factor of two. 
        }
      \end{center} \end{minipage}
  \end{figure*}

%%%%%%%%%%%%%%%%%%%%%%%%%%%%%%%%%%%%%%%%%%%%%%%%%%%%%%%%%%%%%%%%%%%%%%%%%%%%

\subsection{Smart Stars}
As the gas density increases in high density regions, hydro codes, including \enzoc, require a method
to convert the high density gas into stars in many cases. This is done to deal
with gas which has reached the  maximum allowed refinement level of the simulation and for which
further collapse is being artificially suppressed through artificial pressure support.
Gas which has reached this point is expected to result in star formation and so star particles are
introduced. In this work we introduce particles to mimic the formation
of both Pop III stars and SMSs. Upon formation all stars are assumed to be stars with low surface
temperatures that are appropriate for main sequence SMSs and less massive proto-stars on the
Hayashi track. Rapidly accreting (i.e. $\dot{M_*} \gtrsim 0.04$ \msolarc/yr) proto-stars
  carry large amounts of entropy (hot accretion) into the stellar interior. The stellar radius
  monotonically increases as the stellar mass increases obeying an analytic mass-radius relation
 \citep{Hosokawa_2010, Hosokawa_2012, Hosokawa_2013}
  \begin{equation}
    R_* \approx 2.6  \times 10^3 R_{\odot} \Big( \frac{M_*}{100 M_{\odot}}\Big){1/2}
  \end{equation}
  where $R_*$ is the stellar radius and $M_*$ is the stellar mass. Note that the relation
    is independent of the actual accretion rate. The stellar interior remains inhomogeneous and subsequently
    contracts radiating energy away. A surface layer containing a small fraction of the mass inflates
    leading to a puffy SMS with low effective temperatures. The expansion continues until the radius
    eventually begins to contract when the mass of the star exceeds $M_* \gtrsim 3 \times 10^4$
  \msolarc. This
    occurs because H$^-$ bound-free opacity, which keeps the stellar surface temperature locked at close to 5000 K
    becomes unavailable as the density in the surface layer drops below $10^{-11}\ \rm{g\ cm^{-3}}$. Nonetheless,
    the radius at this stage of its evolution is approximately 100 AU. In the case where the accretion rates are
    more sedate and less than a  critical threshold of 0.04 \msolarc/yr \citep{Sakurai_2016} then the
    accretion is referred to as cold accretion. The gas contains much less entropy and the radius does
    not increase monotonically with mass. In this case the stellar evolution is that of a normal Pop III
    star. In the case of sustained high accretion rates our \smartstar remains a SMS. On the otherhand
    if the rates are detected to fall below the critical threshold
then the star automatically converts to a Pop III star reflecting the contraction of the Pop III star
to the main sequence on the Kelvin-Helmholtz timescale.
The type of the star will determine the radiation feedback from the star. Pop III stars are modelled
assuming a blackbody spectrum with an effective temperature of T$_{eff} = 10^5 $K \citep{Schaerer_2002}
while SMSs are modelled by assuming a blackbody spectrum with an effective temperature
of T$_{eff}$ = 5500 K \citep{Hosokawa_2013}. The luminosity rates for the PopIII star as given
  by  \cite{Schaerer_2002} have recently been verified by \cite{Haemmerle_2017b} who recover the
  rates of \cite{Schaerer_2002} for cases where the accretion rate is below the critical rate.
  Nonetheless, including the more recent and updated rates of \cite{Haemmerle_2017b} will be included
in future development of the algorithms.\\
\indent In order to form a \smartstar particle we assess the following criteria on every timestep:
\begin{enumerate}
\item The cell is at the highest refinement level
\item The cell exceeds the Jeans density
\item The flow around the cell is converging along each axis
\item The cooling time of the cell is less than the freefall time
\item The cell is at a local minimum of the gravitational potential
\end{enumerate}
The Jeans density is calculated using the prescription given in \cite{Krumholz_2004} which itself
follows from the Truelove criteria \citep{Truelove_1997, Truelove_1998}.
We calculate the gravitational potential in a region of twice the Jeans length around the cell.
We experimented with also including the additional conditions relating to the gas boundedness and the
Jeans instability test (see \cite{Federrath_2010} for more details). However, we found that these
additional tests were sub-dominant compared to the criteria noted above and so in the interest of
optimisation we did not include them.\\
\indent When calculating the velocity of the particles subsequent to accretion events we explicitly
invoke conservation of momentum to determine the updated velocity of the particle. Doing so
ensures we explicitly conserve linear momentum within the system \citep{Krumholz_2004}. 

\subsection{Accretion onto the \smartstar}
\indent Once a \smartstar is formed it can accrete gas within its accretion radius (4 cells)
and it can merge with other \smartstar particles. Accretion onto the \smartstar is determined by
calculating the flux of gas across the accretion surface.
\begin{equation}
  \dot{M} = 4\pi \int_S {\rho v_r^- r^2 dr}
\end{equation}
where $\dot{M}$ is the mass accretion rate, $S$ is the surface over which we integrate, $\rho$ is the
density of the cells intersecting the surface, $v_r^-$ is the velocity of cells intersecting
the surface and which have negative radial velocities and $r$ is the radius of our surface.
The surface, $S$, is
the surface of a sphere with radius the accretion radius. As noted above we set the accretion
radius to be 4 cells, we choose to fix this radius independent of the resolution or the mass of
the \smartstarc. We do this so as to be as accurate as possible when calculating the accretion
rate, any mass travelling radially inward at a distance of four cells from the \smartstar is taken
to be accreted onto the \smartstar - we therefore strive for the maximum possible physical
resolution. For completeness we experimented with increasing the accretion radius up to 64 cells
and found only small variations in our accretion rate results. Fragmentation is, however, greatly
reduced as \smartstars are prevented from forming within the accretion radius of another \smartstarc.
The accretion onto the star is calculated at each timestep,
however this is likely to be a very noisy metric. To alleviate this to some degree we average
the accretion rate over intervals of 1 kyr and use that averaged accretion rate as the actual accretion
rate. The accretion rate is added as an attribute to each star and hence a full accretion history
of every \smartstar is outputted as part of every snapshot. \\
\indent Mergers with other \smartstars are also included in the accretion onto the \smartstarc. In
this case the more massive \smartstar retains its information (i.e. age, type etc) after the merger
event - information on the less massive \smartstar is lost. The mass of the
less massive \smartstar is added to the accretion rate of the more massive \smartstar for that
timestep. \smartstars are merged when they come within an accretion radius of each other.

%%%%%%%%%%%%%%%%%FIGURE 2%%%%%%%%%%%%%%%%%%%%%%%%%%%%%%%%%%%%%%%%%%%%%%%
\begin{figure*}
  \centering 
  \begin{minipage}{175mm}      \begin{center}
      \centerline{
        \includegraphics[width=9cm]{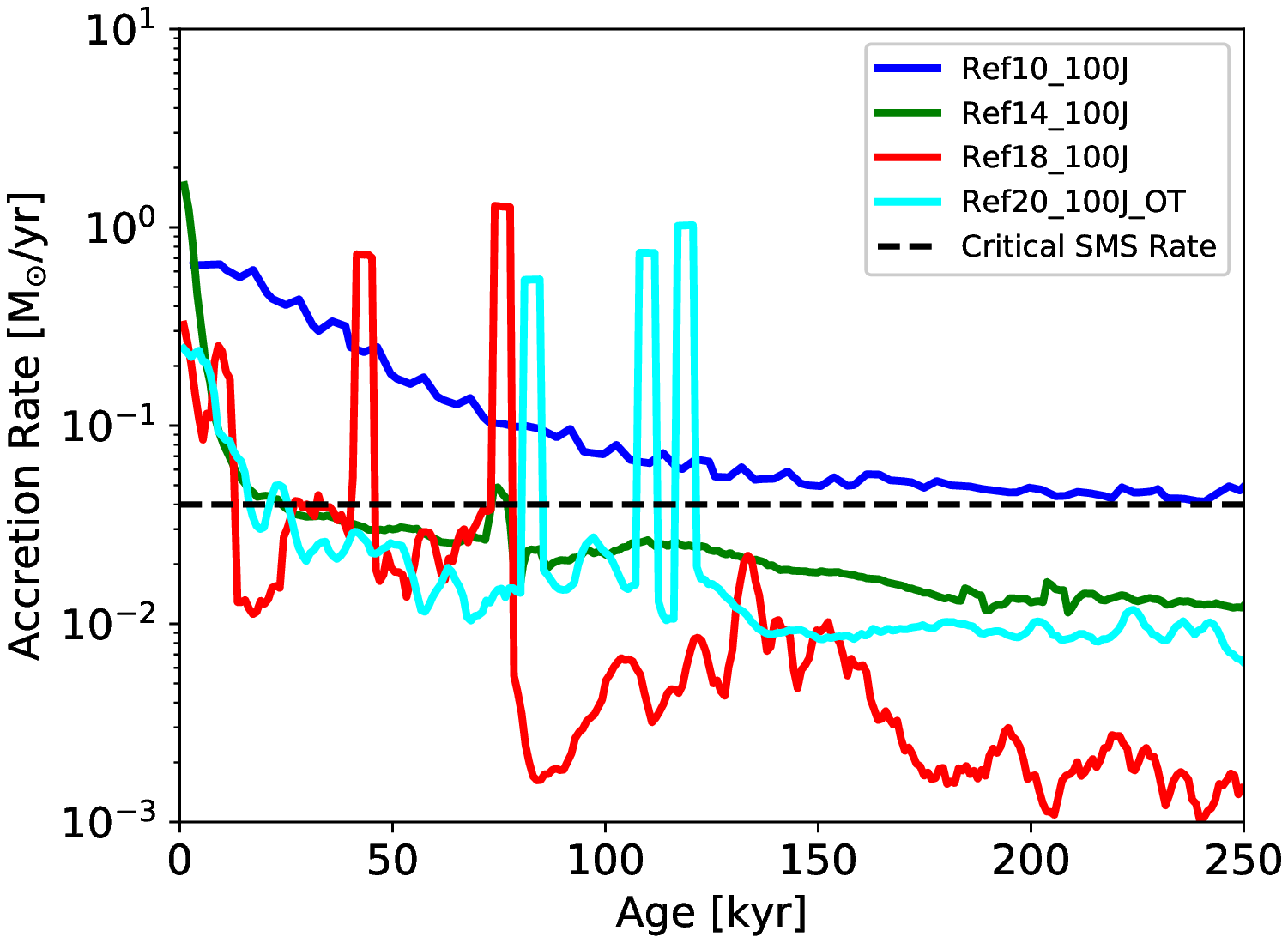}
        \includegraphics[width=9cm]{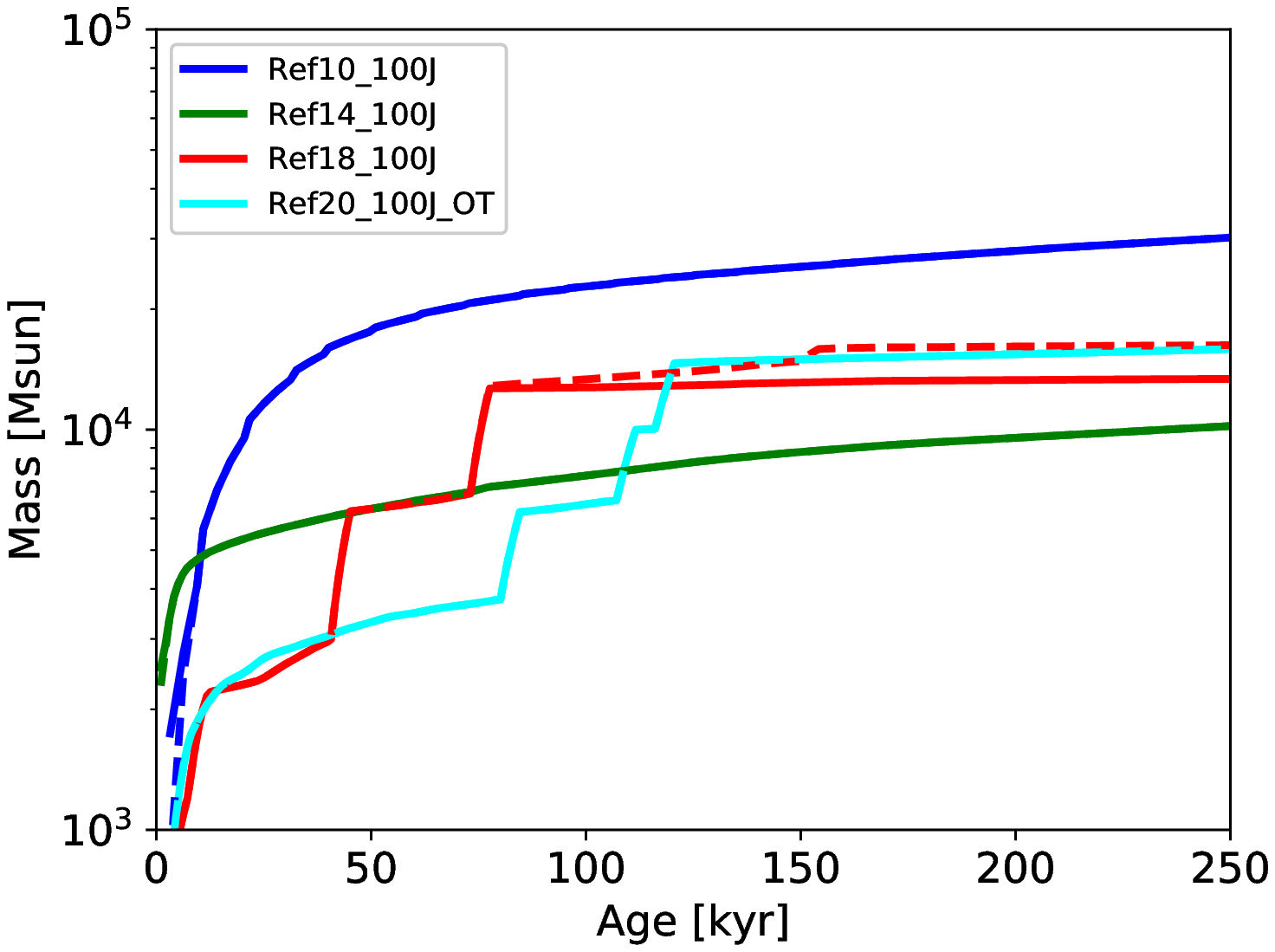}}
        \caption[]
        {\label{MassAccretionRate_Resolution}
          The mass accretion rate onto the largest \smartstar particle in each simulation
          for the 100 J$_{21}$ case. The only difference between each simulation is the
          maximum resolution. As the resolution is increased we see a general trend of decreasing
          accretion rate. There are several factors at play, firstly the increased resolution
          allows for the capture of fragmentation within the collapsing gas with competitive
          accretion ensuing between the (massive) stars. Related to this is the fact that higher
          resolution leads to clumps being resolved and potentially to a more chaotic accretion
          pattern as seen in the Ref18 and Ref20 simulations in particular. This is because as the
          resolution increases we are able to better resolve the small scale structure of the gas
          surrounding our accretion zone. In the right hand panel we show the
          mass of the largest particle in each case as well as the cumulative mass of all of the
          \smartstar particles in each simulation. As expected from the mass accretion rates in
          the left hand panel the Ref10 simulation shows the largest masses. The Ref18 and Ref20
          show excellent convergence after approximately 150 kyr but with differences, due to
          the different accretion histories, before that.
        }
      \end{center} \end{minipage}
  \end{figure*}

%%%%%%%%%%%%%%%%%%%%%%%%%%%%%%%%%%%%%%%%%%%%%%%%%%%%%%%%%%%%%%%%%%%%%%%%%%%%

\subsection{Feedback from the \smartstar} \label{feedback}
Upon formation a \smartstar is initially assigned a SMS type. The type of the star is only
determined by its mass accretion rate and is independent of the chemical environment. The \molH
fraction is not a criteria for forming stars. In Pop III studies, Pop III stars are generally only formed
when the \molH fraction is greater than $10^{-4}$ \citep[e.g.][]{OShea_2007b}. However, since we
are primarily concerned with SMS formation in a metal-free context our criteria must be different.
If the accretion rate onto the star falls below the critical threshold of 0.04 \msolarc/yr then
the star changes type and becomes a Pop III star \citep{Sakurai_2016}. The type of the star determines its feedback.
Pop III stars are modelled assuming a blackbody spectrum with a characteristic mass of 40 \msolarc
 (Table 4, \cite{Schaerer_2002}). From that we assign a \textit{LuminosityPerSolarMass} to the
Pop III star and the star consequently becomes more luminous and the feedback more intense as the mass
of the star increases. We allow accretion onto the star in tandem with feedback. Since the
spectrum of a Pop III is strong in the UV it will strongly ionise the surrounding medium. However,
in regions of very high infall the HII regions of the Pop III star are unable to expand
more than a few cells from the star due to the extreme opacity of the surrounding gas \citep{Chon_2018}. \\
\indent  SMSs are modelled by assuming
a blackbody spectrum with an effective temperature of T$_{eff}$ = 5500 K \citep{Hosokawa_2013}.
The radiation spectrum for a SMS therefore peaks in the infrared as opposed to the UV for Pop III stars.
For the specific luminosity of the SMS we take a characteristic mass of 500 \msolar and apply
the contribution from the non-ionising photons only \citep{Schaerer_2002}. As with the 'normal'
Pop III stars the SMS luminosity changes as mass is accreted and the total luminosity then
scales up as the mass increases. \\
\indent In both cases the radiation from the stars is propagated outwards from the star using the
\texttt{MORAY} radiative transfer package \citep{WiseAbel_2011} that is part of \enzoc.
\texttt{MORAY} is able to model the ionisation of H, He and He$^{+}$. It can also account for the
photo-dissociation of \molH for photons with energies within the Lyman-Werner band and the
photo-detachment of  $\rm{H^-}$ and $\rm{H_2^+}$ for photons in the infrared band. For each type of
star we use five energy bins. The first two energy bins (E $< 13.6$ eV) are weighted by the cross
section peaks for $\rm{H^-}$,  $\rm{H_2^+}$ and \molH photo detachment/dissociation respectively.
The next three energy bins are determined using the \texttt{sedop} code developed by
\cite{Mirocha_2012} which determines the optimum number of energy bins needed to
accurately model radiation with energy above the ionisation threshold of hydrogen. For the
self-shielding of \molH against LW radiation we use the prescription of \cite{Wolcott-Green_2011}.\\
\indent \texttt{Moray} is also able to switch to an optically thin mode of radiative transfer
on demand. For the highest resolution runs this can be especially useful due to the fact that
propagating photons through a high density AMR mesh becomes computationally prohibitive. For those
cases we include only the feedback in the Infrared and LW bands - which are of most relevance to
SMSs and we assume that the radiation from Pop III stars is effectively halted close to the star
(which is a reasonable assumption at least in the first 100 kyr \citep{Chon_2018}).

%%%%%%%%%%%%%%%%%%%%%%%%%%%%%%%%%%%%%%%%%%%%%%%%%%%%%%%%%%%%%%%%%%%%%%%%%%%%
%%%%%%%%%%%%%%%%%FIGURE 3%%%%%%%%%%%%%%%%%%%%%%%%%%%%%%%%%%%%%%%%%%%%%%%
\begin{figure}
  \centering 
  \includegraphics[width=0.47\textwidth]{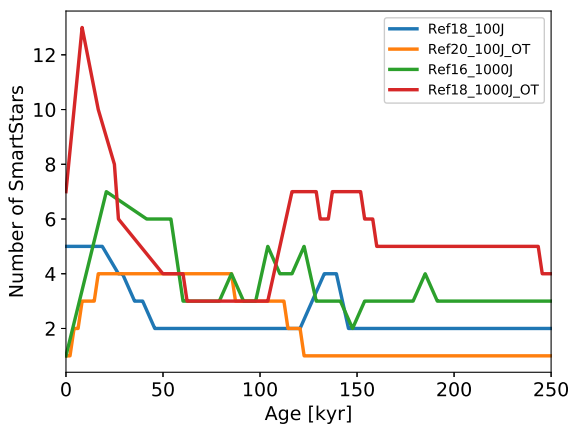}
    \caption{
      The number of \smartstars as a function of time for the highest resolution
      runs for the case of the 100  J$_{21}$ and 1000  J$_{21}$ backgrounds. In both
      cases the level of fragmentation peaks at the start and due to mergers and
      accretion saturates after less than 200 kyr. The number of \smartstars is
      systematically higher for the higher background due to the fact that the
      central gas core, out of which the \smartstars form is larger for the
      higher background. 
    }
    \label{Fragments}
\end{figure}

%%%%%%%%%%%%%%%%%%%%%%%%%%%%%%%%%%%%%%%%%%%%%%%%%%%%%%%%%%%%%%%%%%%%%%%%%%%%
%%%%%%%%%%%%%%%%%FIGURE 4%%%%%%%%%%%%%%%%%%%%%%%%%%%%%%%%%%%%%%%%%%%%%%%
\begin{figure*}
  \centering 
  \begin{minipage}{175mm}      \begin{center}
      \centerline{
        \includegraphics[width=18cm]{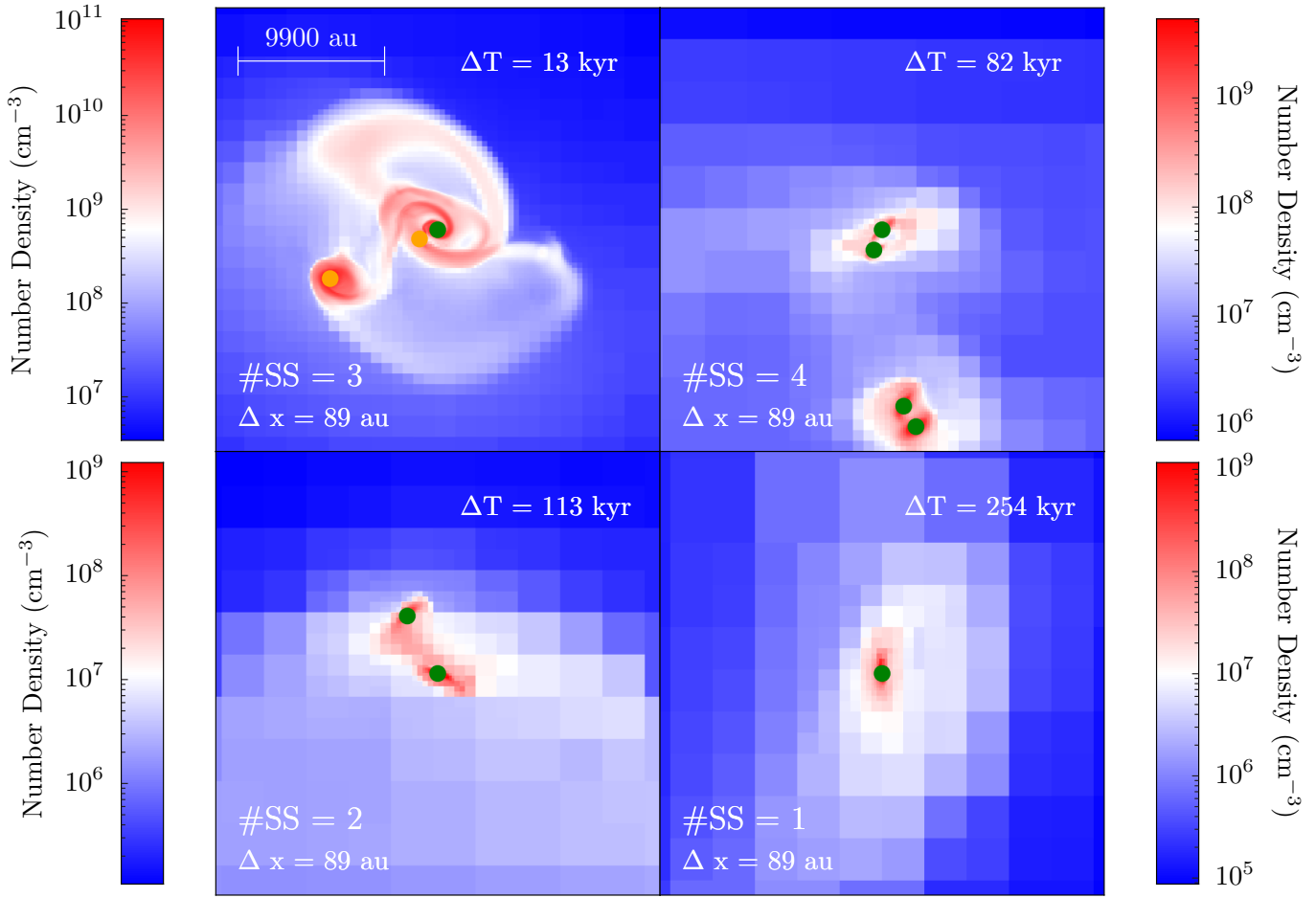}}
        \caption[]
                {\label{Ref20Fragments}
                  A projection along the ``y'' axis of the box centred on the most massive
                  \smartstar in each panel. The depth of each projection is 30000 AU and is
                  density weighted. Shown is the 
          evolution of the clumps within the collapsing halo for the Ref20\_100J\_OT
          simulation. Green circles indicate Pop III stars while orange circles indicate SMSs.
          Initially two clumps form at the centre of the collapsing
          structure. The gas continues to fragment forming two separate binary systems.
          The double binary evolves before the individual clumps within the binary merge
          - the first after 80 kyr and the second subsequently after 110 kyr.
          The binary, that is left, quickly merges then to form a single Pop III star
          at the centre of the halo with no further fragmentation observed up to 250 kyr. No three
          body interactions occur in this simulation.
        }
      \end{center} \end{minipage}
  \end{figure*}

%%%%%%%%%%%%%%%%%%%%%%%%%%%%%%%%%%%%%%%%%%%%%%%%%%%%%%%%%%%%%%%%%%%%%%%%%%%%
%%%%%%%%%%%%%%%%%FIGURE 5%%%%%%%%%%%%%%%%%%%%%%%%%%%%%%%%%%%%%%%%%%%%%%%
\begin{figure*}
  \centering 
  \begin{minipage}{175mm}      \begin{center}
      \centerline{
        \includegraphics[width=18cm]{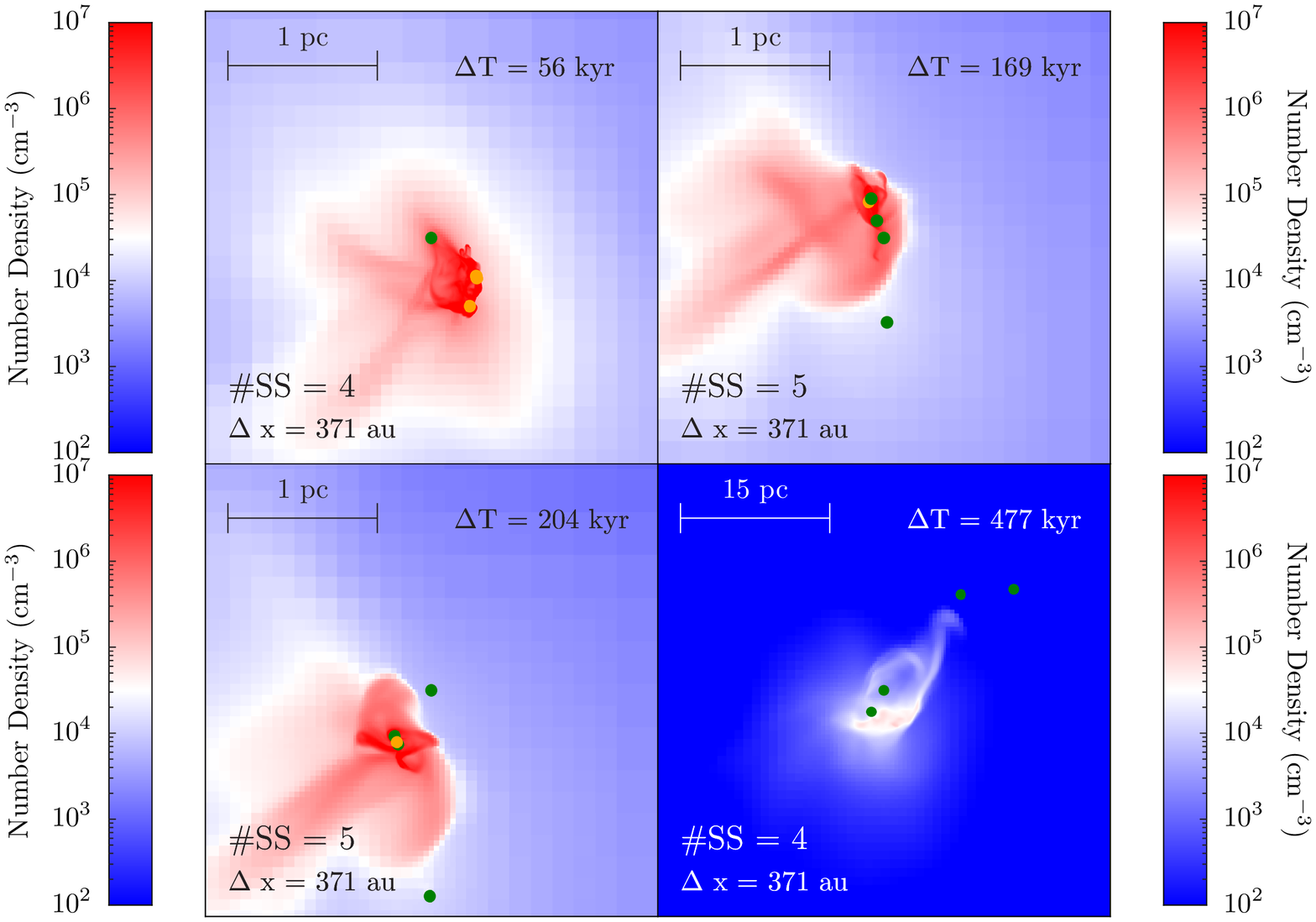}}
        \caption[]
          {\label{Ref18Fragments}
            We show visualisations of the Ref18\_1000J\_OT simulation for four different
            times extending up to nearly 500 kyr after the formation of the first \smartstarc.
            The visualisations are centred on one of the least massive \smartstars which
            always sits near the centre of the potential. We do this to illustrate the
            ejection of two of the most massive \smartstars from the centre.
          Within the 1000 J$_{21}$ simulation 4 \smartstars exist after 56 kyr.
          As the simulation progresses one additional \smartstar forms and then merges
          again with the central most massive SMS. Between 100 kyr and 200 kyr several
          multiple body interactions combined with the depleted central gas densities
          cause the two most massive \smartstars to become ejected (see bottom right panel).
          After approximately 450 kyr (bottom right panel) the
          high density gas has been removed by the accreting stars. SMSs are colour coded
          orange and Pop III stars are coloured green.
        }
      \end{center} \end{minipage}
  \end{figure*}

%%%%%%%%%%%%%%%%%%%%%%%%%%%%%%%%%%%%%%%%%%%%%%%%%%%%%%%%%%%%%%%%%%%%%%%%%%%%

\section{Results} \label{Sec:Results}
In Table 1 we detail the simulations carried out in this study. We explore two background
cases, one with a LW background of 100 J$_{21}$ and one with a background level of 1000 
J$_{21}$. Furthermore, we run one simulation with a background of 1 J$_{21}$ for completeness
but this low level background is unlikely to foster the formation of a SMS.
A background with a value of  1000 J$_{21}$ is likely required to drive a fully
isothermal collapse where \molH is fully suppressed throughout the halo. The 100 J$_{21}$
case suppresses \molH until the core can sufficiently self-shield but regions of the Universe
exposed to 100 J$_{21}$ are likely to be much more common compared to 1000 J$_{21}$
\citep{Ahn_2009, Inayoshi_2015b} and perhaps offer a more realistic condition for forming the vast
majority of SMBH seeds. We begin by examining the accretion rates onto the
\smartstar particles during their evolution.
\subsection{Accretion Histories}
In the left hand panel of Figure \ref{MassAccretionRate} we show the
mass accretion history for the largest \smartstar particle for four different realisations. We use
three different backgrounds 1 J$_{21}$, 100 J$_{21}$ \& 1000 J$_{21}$. For each background we show
the results from simulations at a maximum refinement level of 18 which corresponds
to a spatial resolution of 0.001 pc at z = 25.5 ($\sim 200$ AU). At this resolution we are within an
order of magnitude of the accretion disk radius of a SMS. We could push our resolution to higher values
but instead we focus on extending the runtimes to a significant fraction of the lifetime of the SMS
while also testing for convergence. For the realisation with a background of 1000 J$_{21}$ the
computational expense of the run meant that we compromised by using the optically thin (OT)
approximation for the radiative transfer as described in \S \ref{feedback}. Simulations run
  with the optically thin approximation are suffixed with ``OT'' (e.g. Ref18\_1000J21\_OT
  in Table 1). We therefore also show the results of a simulation run a refinement level of 16
but with full radiative transfer for comparison.
The blue line, representing the accretion history of the \smartstar for the 1 J$_{21}$ background
has the lowest accretion rates and the accretion rate quickly drops below the critical threshold
(dashed horizontal line) - this particle switches type to a Pop III star as expected for a low
background. For the 100 J$_{21}$ (green line) the accretion history shows that the star initially
fluctuates between being a SMS and being a Pop III star for approximately the first 80 kyr of
its lifetime after which the accretion rate recedes below the critical threshold and while there
are subsequent increases in the accretion rate, the rate never again exceeds the critical level.
The red line and the cyan line are the accretion histories for the 1000 J$_{21}$ background. The
difference in the two cases is that for the red line the full radiative transfer is used but the
refinement level is reduced by a factor of four. For the maximum refinement case the radiation
is modelled in the optically thin mode with only non-ionising radiation included. Nonetheless,
the agreement is remarkable. The high accretion rates mean that the star is always of type SMS and
so the radiation is confined to the non-ionising regime. Even more satisfying is the
convergence between the simulations. The slightly lower resolution run (red line) gives broadly the
same accretion history as the higher resolution run (cyan line). \\
\indent In the right hand panel of Figure \ref{MassAccretionRate} we plot the total mass accreted
by the largest particle in each simulation. We also overplot the total stellar mass accreted by all
particles in each simulation (dashed lines). Starting with the 1 J$_{21}$ background we see that
the accretion rates lead to a total mass of a few hundred \msolar after 250 kyr at which point the
accretion rates have dropped significantly and this will be close to the final mass. While this mass
is clearly above the characteristic mass found by other authors \citep{Turk_2009, Stacy_2010,
  Clark_2011a, Stacy_2012} our resolution is too low to resolve fragmentation in these Pop III systems
and the mass ascribed here should be taken as a reasonable  upper limit. For the Ref18\_100J simulation
(green line), the mass
of the largest \smartstar is closer to $10^4$ \msolar while for the 1000 J$_{21}$  the mass
has increased by an order of magnitude up to $10^5$ \msolarc. The difference between the red and cyan
line is this case again emphasises the excellent convergence that is achieved with a difference in
the final mass (after 250 kyr) of less than a factor of 2. \\
\indent In Figure \ref{MassAccretionRate_Resolution} we again test for convergence in our spatial
resolution using the 100  J$_{21}$  background simulations only. We look at the difference between
maximum spatial resolutions that range from 10 levels of refinement up to 20 levels of refinement
(corresponding to $\sim$ 10000 AU down to $\sim$ 90 AU). The very high resolution Ref20 simulation uses the
optically thin approximation for the non-ionising radiation and neglects ionising photons.
The blue line,
represents the accretion history of the Ref10 simulation and has the highest accretion rate 
over the course of the first 250 kyr and exceeds the critical rate. The accretion rate then
drops below the critical rate for the remainder of the lifetime of the star. Both the Ref14, Ref18 and
the very high resolution Ref20 simulations show lower accretion rates. The Ref18 and Ref20
simulations show more variation compared to the Ref14 simulation due to the increased resolution and
therefore the ability of the hydro solver to identify smaller clumps of material which may then
subsequently be accreted by the central core. However, the degree of convergence between the
Ref20 (cyan) and Ref18 (red) lines is again quite remarkable. There is more initial fragmentation in
the Ref18 simulation and then those fragments (stars) merge (this is the main reason behind the spikes in accretion
seen at 45 and 70 kyr). \\
\indent In the right hand panel of Figure  \ref{MassAccretionRate_Resolution} we show the total stellar
mass for both the largest \smartstar and for the total mass from all the \smartstars formed. The
lowest resolution, blue line, clearly overestimates the accretion rate and hence the final mass and
this has been shown before to be a problem for low resolution simulations \citep[e.g.][]{Negri_2017}.
The maximum resolution for the Ref10 run is approximately 0.44 pc, the bondi radius
of a star with a mass of 1000 \msolar in an atomic cooling halo is of the order of 0.5 pc.
Given the Ref10 simulation is not able to resolve the Bondi radius around the accreting star even
once it's mass reaches 1000 \msolar it is not surprising that the mass is overestimated. Only once the
refinement is sufficient to resolve the bondi radius do we begin to see convergence. After 250 kyr
the difference in total stellar mass between the Ref14, Ref18 and Ref20 runs is less than a factor of
2. The total stellar mass converges towards a mass of approximately $10^4 $
\msolar after 250 kyr. However, the largest portion of the growth of the \smartstars takes place
within the first 100000 years. The sharp spikes seen in the Ref18 and Ref20 runs indicate the
somewhat chaotic nature of the accretion. 
No fragmentation (see below) is seen for the Ref14 simulation (hence no dashed line) while
multiple fragments (stars) are found for both the Ref18 and Ref20 simulations. We will now examine the
fragmentation in our simulations in more detail. 
%%%%%%%%%%%%%%%%%FIGURE 6%%%%%%%%%%%%%%%%%%%%%%%%%%%%%%%%%%%%%%%%%%%%%%%
\begin{figure}
  \centering
  \includegraphics[width=0.47\textwidth]{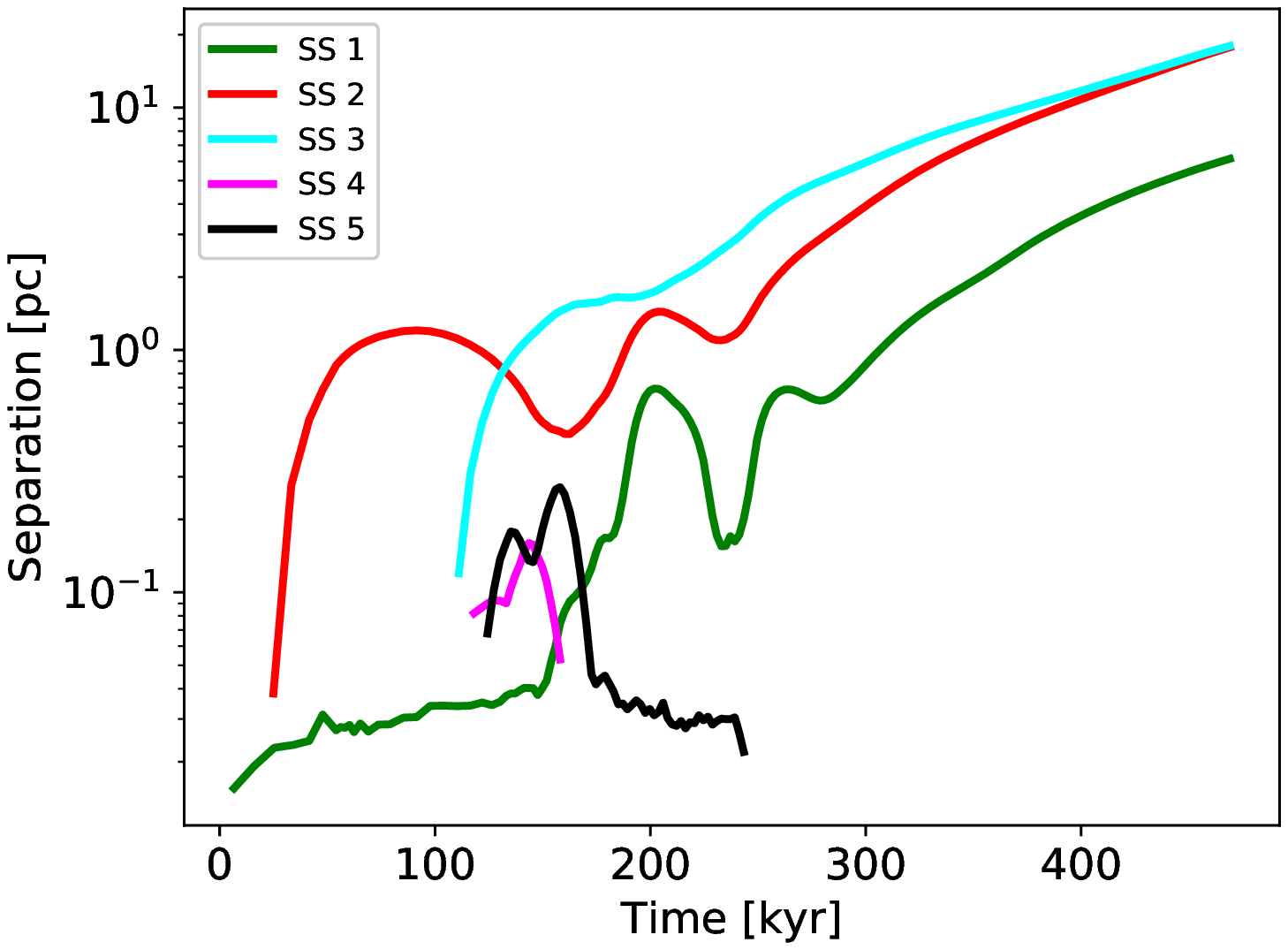}
  \includegraphics[width=0.47\textwidth]{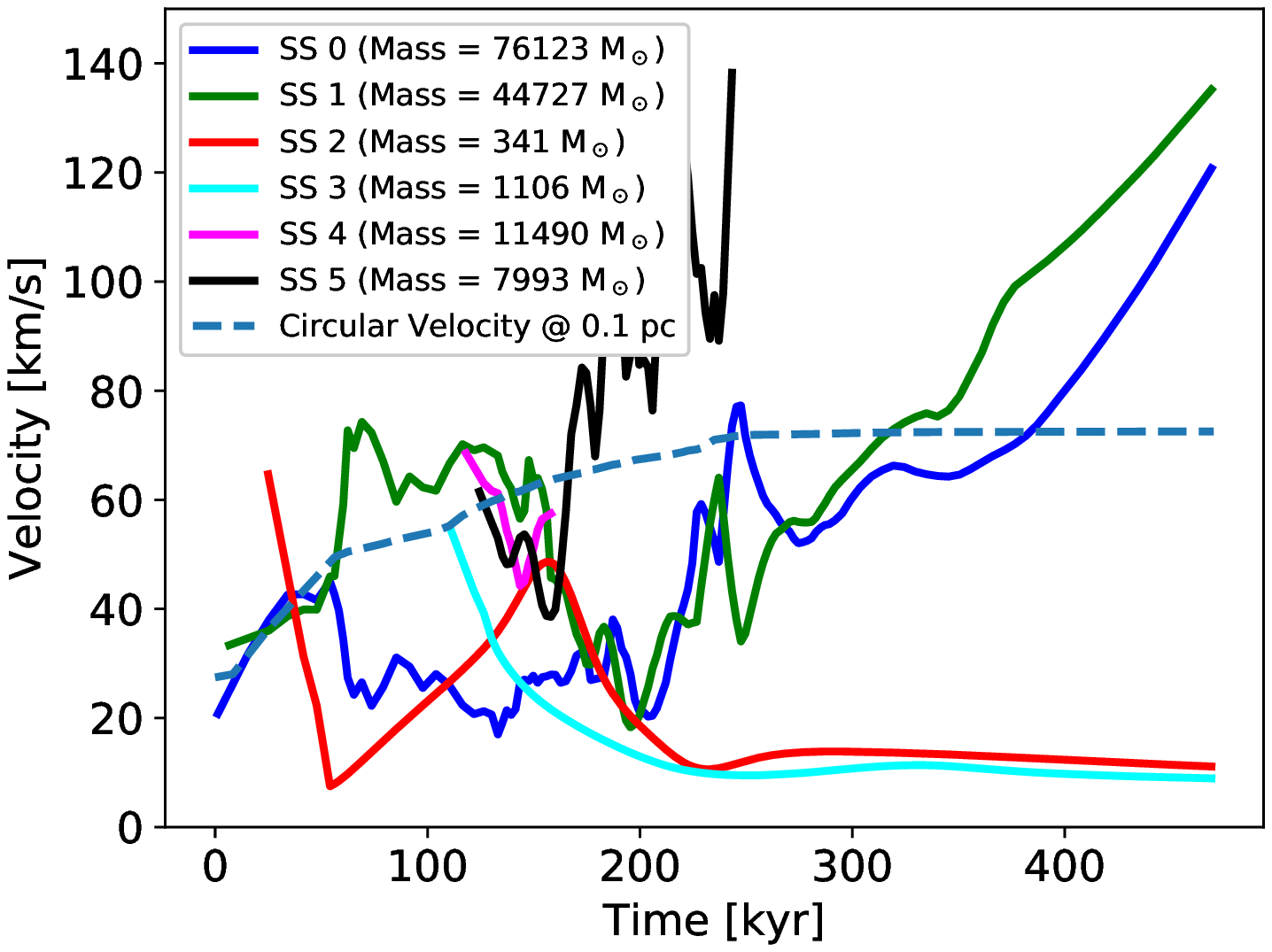}
    \caption[]
            {\label{FragmentVelocities}
              The evolution of the separations between each \smartstar and the most massive
              \smartstar and the velocities of all of the \smartstars in the Ref18\_1000J\_OT
              simulation. Only \smartstars which survive for more than 10 kyr are plotted.
              \textit{Upper Panel:} The separation between the most massive \smartstarc, SS 0, and
              the other \smartstarsc. All of the \smartstars are born, with the possible exception
              of SS 3, within 0.1 pc of the most massive \smartstarc. As the system develops and
              multiple three body interactions occur SS 0 and SS 1 are ejected from the halo centre.
              SS 4 and SS 5 are involved in mergers at 150 kyr and 250 kyr respectively. 
              \textit{Bottom Panel:} The evolution of the velocities of the \smartstarsc.
              The circular velocity of the
              halo within 0.1 pc of the centre of the halo is also shown. Several
              multi-body interactions between 100 kyr and 200 kyr result in the
              ejection of SS 0 and SS 1 (their masses are 76123 \msolar and
              44727 \msolar respectively). Two lower mass \smartstars are left at the
              centre of the halo with low velocities. 
            }
\end{figure}

%%%%%%%%%%%%%%%%%%%%%%%%%%%%%%%%%%%%%%%%%%%%%%%%%%%%%%%%%%%%%%%%%%%%%%%%%%%%
%%%%%%%%%%%%%%%%%FIGURE 7%%%%%%%%%%%%%%%%%%%%%%%%%%%%%%%%%%%%%%%%%%%%%%%
\begin{figure*}
  \centering 
  \begin{minipage}{175mm}      \begin{center}
      \centerline{
      \includegraphics[width=9cm]{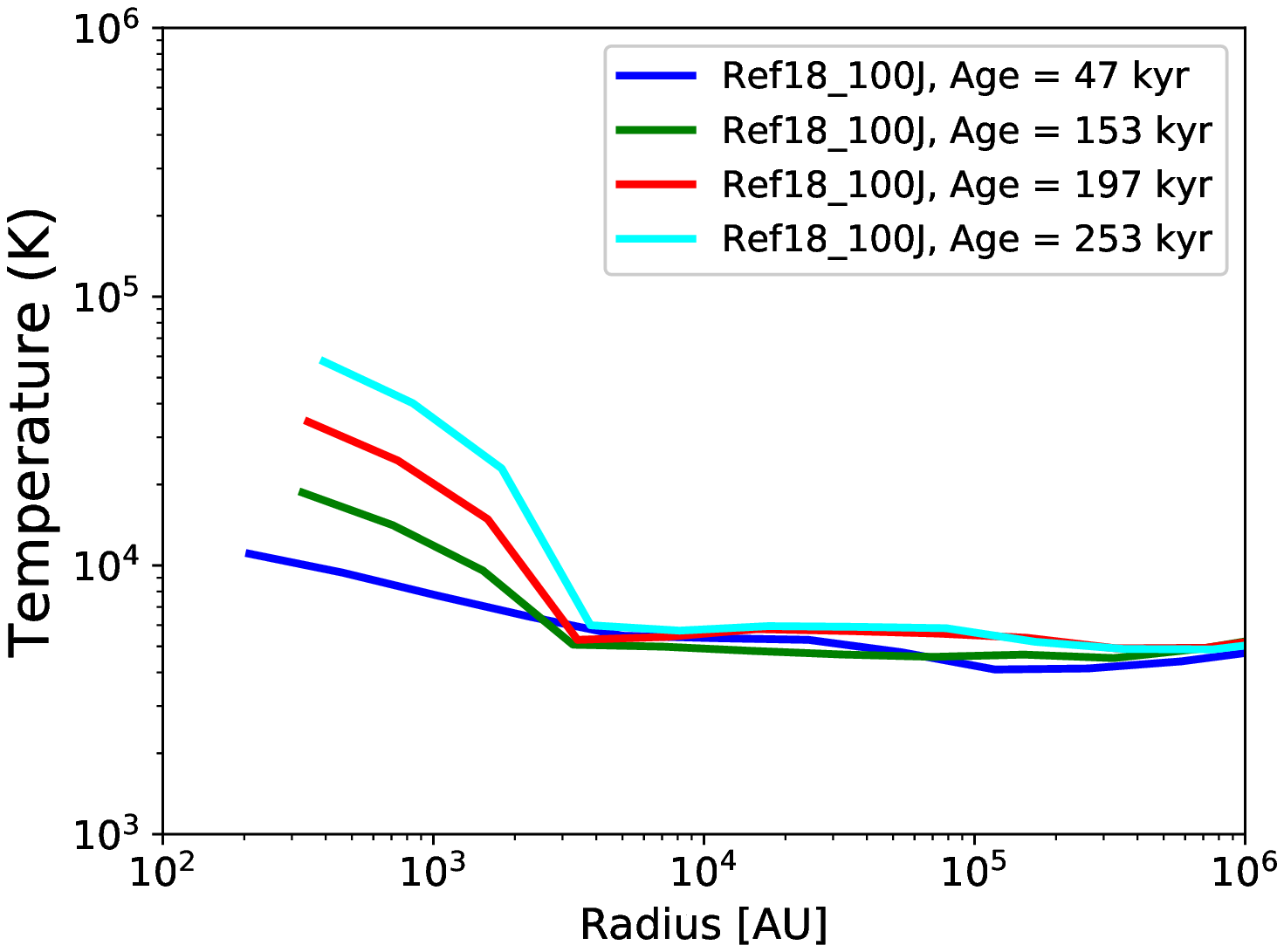}
      \includegraphics[width=9cm]{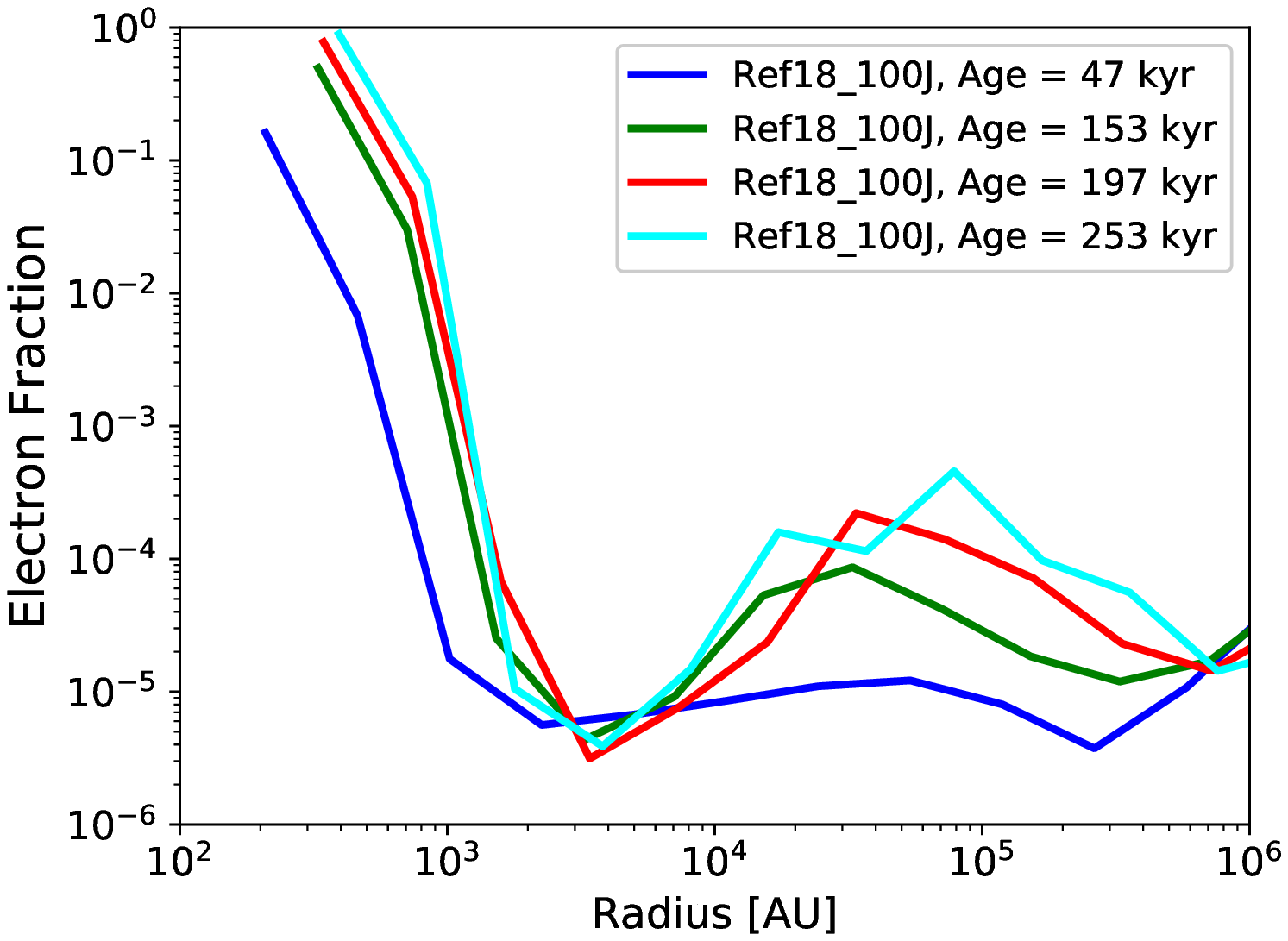}}
        \caption[]
        {\label{MultiPlot}
          Radial profiles of the Ref18\_100J simulation for 4 different
          times. The star initially fluctuates between being a SMS and a
          Pop III before the accretion rate drops and the star contracts
          to the main sequence. After 50 kyr the star is always in the
          Pop III cycle. In the left hand panel we show the temperature
          profile and in the right hand panel the electron fraction.
          All profiles are centred
          on the most massive star in the simulation. The maximum resolution
          of the Ref18 simulation is approximately 200 AU. The high gas density
          surrounding the star means that radiative cooling is still very effective
          at regulating the temperature as well as containing the HII region.
          The electron fraction, while very high close to the star, drops sharply
          only a few cells from the star.
        }
  \end{center} \end{minipage}       
\end{figure*}

%%%%%%%%%%%%%%%%%%%%%%%%%%%%%%%%%%%%%%%%%%%%%%%%%%%%%%%%%%%%%%%%%%%%%%%%%%%%
%%%%%%%%%%%%%%%%%FIGURE 8%%%%%%%%%%%%%%%%%%%%%%%%%%%%%%%%%%%%%%%%%%%%%%%
\begin{figure*}
  \centering 
  \begin{minipage}{175mm}      \begin{center}
      \centerline{
        \includegraphics[width=18cm]{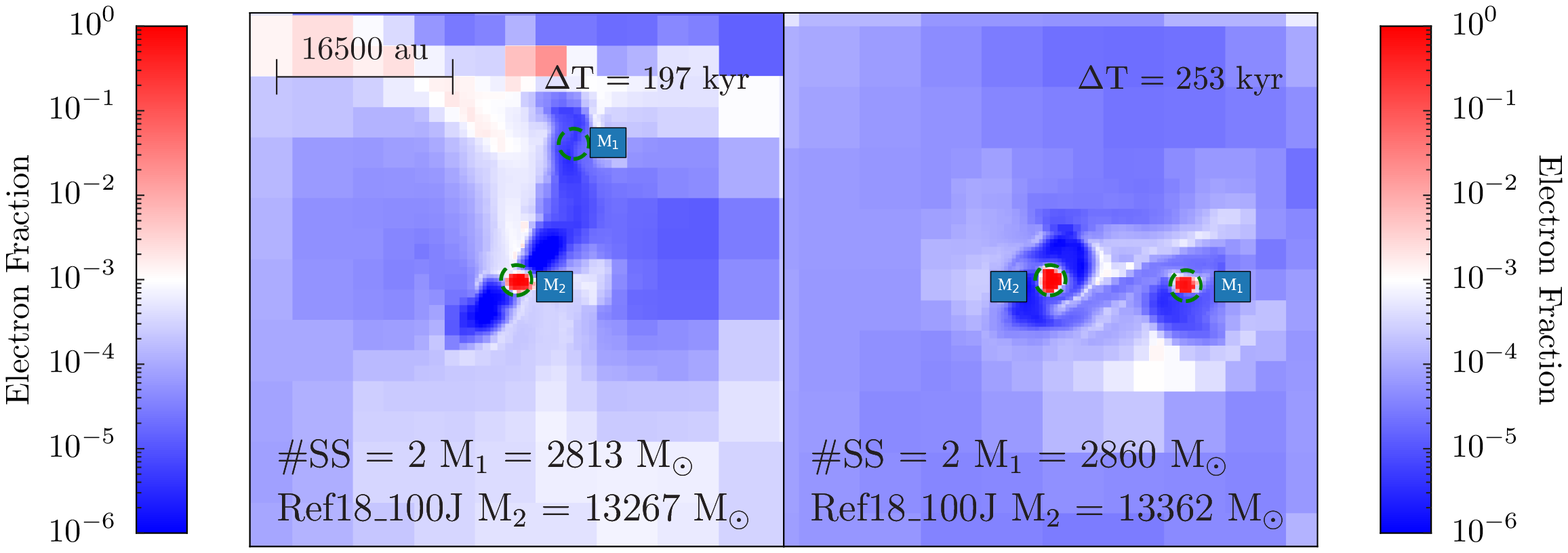}}
        \caption[]
        {\label{TwoPanel}
          Slices of the Electron Fraction for the last two output times
          shown in Figure \ref{MultiPlot}. The dashed green lines identify the Pop III in each case
          and are set to the radius of the accretion sphere. The electron fraction observed
          around the central, most massive star indicates the extent of the HII
          region and that it is confined to a volume close to the star.
        }
      \end{center} \end{minipage}
  \end{figure*}

%%%%%%%%%%%%%%%%%%%%%%%%%%%%%%%%%%%%%%%%%%%%%%%%%%%%%%%%%%%%%%%%%%%%%%%%%%%%

\subsection{Fragmentation within the star forming galaxy}
\indent In Figure \ref{MassAccretionRate_Resolution} neither the Ref10 or the Ref14 runs show any
signatures of fragmentation. In both cases only a single central star forms. However, as we
increase the resolution we are able to better capture fragmentation, especially for the 1000
J$_{21}$ backgrounds where the accretion rates and hence masses of the stars are larger.
In Figure \ref{Fragments} we plot the number of \smartstars
as a function of time. Time equal to zero corresponds to the formation of the first \smartstar in the
simulation. We plot the evolution of the two highest resolution runs in the 100 J$_{21}$ and
1000 J$_{21}$ cases. In all cases there is initially a spike in fragmentation - most pronounced in the
1000 J$_{21}$ due to the more massive halo under going collapse. In the case of the Ref18\_100J
simulation five \smartstars form within 7 kyr of the first \smartstar in a small clustered region
with an average separation of approximately 2500 AU between the \smartstarsc. Similarly, the
Ref18\_1000J undergoes initial vigorous fragmentation with seven \smartstars forming within 5.5 kyr
of each other again in a clustered region with an average separation again of approximately 2500 AU.
Recall, that the merger and accretion radii of the \smartstars in the Ref18 simulations is four cell
widths corresponding to approximately 1500 AU. The stars are therefore forming just outside the merger
radius of the stars. As a result in all cases the
\smartstars tend to merge and the system reaches an equilibrium after almost 150 kyr. The 100  J$_{21}$
case is especially interesting. After 50 kyr the Ref20 run has 4 \smartstars while the
lower resolution Ref18 run has two \smartstarsc. Upon closer inspection (see Figure
\ref{Ref20Fragments}) we see that the Ref20 system is made up of two separate binaries. The Ref18
run was simply unable to resolve the fragmentation of the separate systems into binaries. Similarly,
the 1000 J$_{21}$ simulations show that the higher resolution systems are able to resolve more
fragments. \\
\indent In Figure \ref{Ref20Fragments} we show 
visualisations at 4 different epochs showing the emergence of fragments (stars) during the evolution
of the simulation with a background of 100 J$_{21}$ and at the maximum refinement level.
The projection is aligned along the ``y'' axis of the simulation centred on the most
  massive \smartstar in each panel. The depth of each projection is 30000 AU and is density weighted.
  In the top left panel the simulation axis is fortuitously aligned with the spiral arms of
  the accretion disk. Any accretion disk is however short-lived as the motions of both the gas and the
  \smartstars destroy any coherent disks and deplete the gas reservoir.
In the top left panel we show the initial state of the system 7 kyr after the formation
of the first \smartstar particle. At this stage two particles have formed. As the system evolves
another binary system develops at a separation of approximately 10000 AU. The two binary systems
remain stable through out the first 100 kyr of their existence before merging (at which point the
simulation results converges to the lower resolution Ref18 results). We now examine more closely the
double binary system and the evolution of the system.

%%%%%%%%%%%%%%%%%FIGURE 9%%%%%%%%%%%%%%%%%%%%%%%%%%%%%%%%%%%%%%%%%%%%%%%
\begin{figure}
  \centering
    \includegraphics[width=0.47\textwidth]{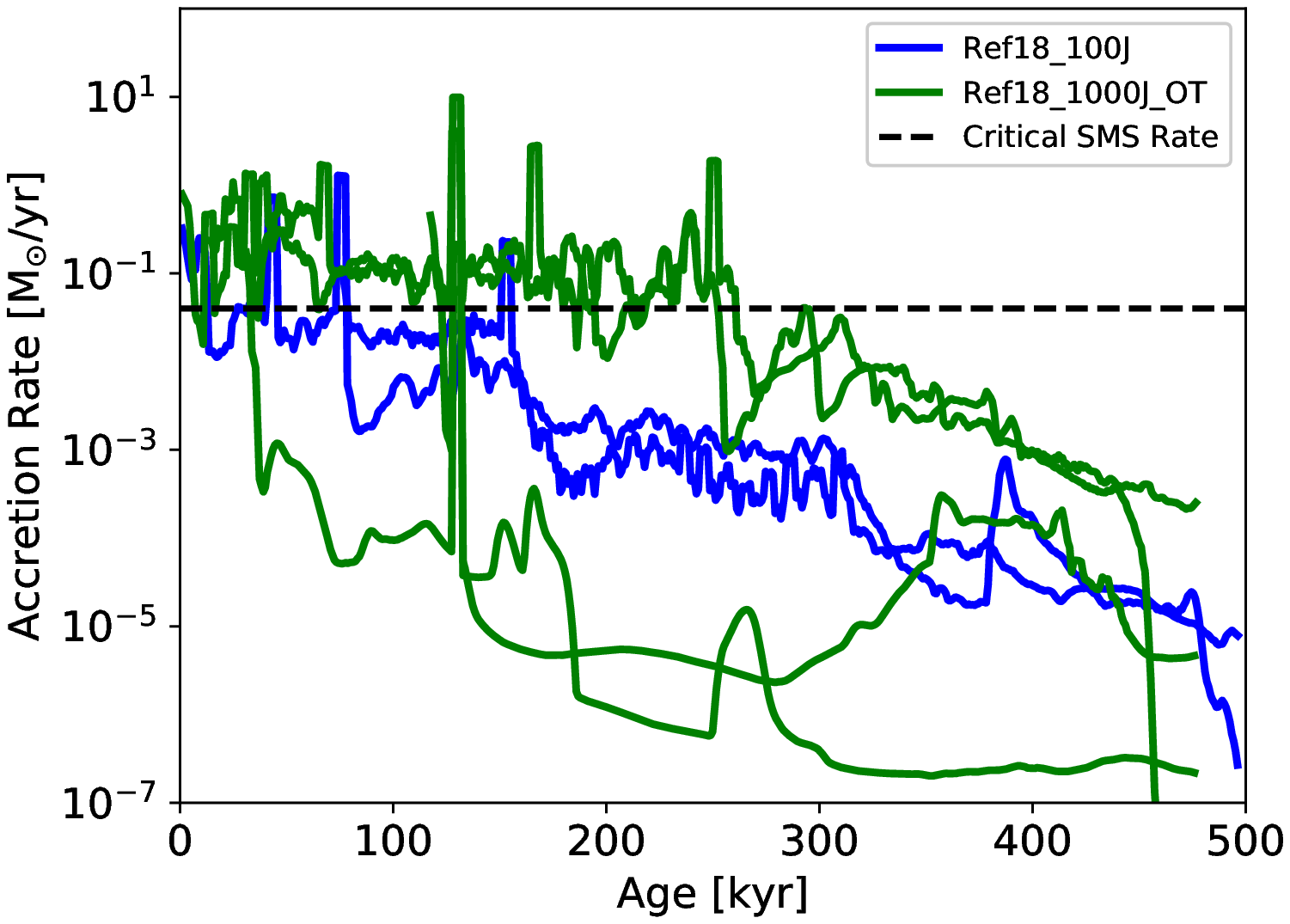}
    \caption[]
            {\label{MassAccretionRate_All} The mass accretion rate onto all
              the \smartstars which survive to the end of the Ref18\_100J and
              Ref18\_1000J\_OT simulations. The Ref18\_100J are coloured blue while
              the  Ref18\_1000J\_OT simulations are coloured green. Two of the
              most massive particles are ejected from the Ref18\_1000J\_OT
              and they show a rapid decrease in accretion rates after the
              ejection (at $\sim 200$ kyr). The other two \smartstars in the
              Ref18\_1000J\_OT have low accretion rates throughout. In the
              Ref18\_100J simulations multi-body interactions early in the
              evolution of the system (before 50 kyr) give large kicks to the
              remaining two particles causing them to be ejected from the
              central core of the halo and causing a decrease in their
              accretion rates. 
            }
\end{figure}

%%%%%%%%%%%%%%%%%%%%%%%%%%%%%%%%%%%%%%%%%%%%%%%%%%%%%%%%%%%%%%%%%%%%%%%%%%%%
\subsection{Binary Systems and Stellar Ejections}\label{ejections}
We select the pair of binary systems that form in the highest
resolution simulation with the 100  J$_{21}$ background. The system is visualised in
Figure \ref{Ref20Fragments}, the stars are colour coded with SMSs coloured orange and Pop III stars
coloured green. All four stars contract to the main sequence and form massive Pop III
stars as the systems evolve. Note that for this very high resolution run we had to switch off
ray tracing and so we do not follow the ionising radiation.
By the time the oldest Pop III star is approximately 250 kyr old it
has a mass of almost $10^4$ \msolarc. The separations' of both binary systems vary between 1000 AU and
3000 AU over the course of almost 100 kyr. After approximately 80 kyr the first binary system coalesces
with the second binary coalescing shortly afterwards. This results in the formation of a ``new'' binary
with an initial separation of approximately 10000 AU. This new binary coalesces after a further 10
kyr leaving a single \smartstar with a mass of $\sim$ 15000 \msolarc. The single \smartstar continues to
accrete mass at a rate of $\dot{M} \sim 10^{-2}$ \msolarc/yr at the centre of the potential.\\
%If any of the  binaries had remained stable beyond the lifetime
%of the massive Pop III stars then this system may have formed a massive black hole binary system
%which would, if the black holes subsequently merged, provide a strong candidate for detection
%with LISA \citep[e.g.][]{Sesana_2007}.\\
\indent The 1000 J$_{21}$ background simulations have a larger number of \smartstars and form a more
diverse system with no tight binaries forming. Due to the very high initial accretion
rates (see Figure \ref{MassAccretionRate}) the mass of the largest \smartstar increases rapidly and
after 250 kyr has a mass of almost $10^5$ \msolarc. In Figure \ref{Ref18Fragments} we show the
evolution of that system at four different times. The number of \smartstars varies between 4 and 5
during the course of the evolution. A series of close encounters between the \smartstars occuring
between 100 kyr and 200 kyr results in the ejection of two of the most massive \smartstars from
the centre of the halo with velocities which exceed the escape velocity of the core of the halo.
In Figure \ref{FragmentVelocities} we plot the separation (upper panel) of each \smartstar from
the most massive \smartstar and the individual particle velocities (bottom panel) as a
function of time. The circular velocity for the halo at a radius of 0.1 pc ($\sim 20000 $AU) is
also plotted. SS 0 is not plotted in the upper panel of  Figure \ref{FragmentVelocities} because the
separations are relative to SS 0. SS1, SS 3, SS 4 \& SS 5 are all in close proximity to SS 0 at
around the 100 kyr mark. Between 100 kyr and 250 kyr the particles all undergo complex dynamical
interactions which lead to mergers and kicks to the particles eventually leading SS 0 to be ejected
from the halo. In the bottom panel of Figure \ref{FragmentVelocities} we plot the velocities of the
particles. The close encounters and mergers (for example star
SS 5 merges with SS 0 after 230 kyr) results in sharp increase to the velocities to a number of
the stars. These kicks are sufficient to eject the two most massive stars from the centre of the
halo with velocities of more than 100 \kms. While the ejected stars initially receive large
accelerations the accelerations quickly dissipate but the velocities and trajectories at this point
are more than sufficient to escape the halo. The two
least massive stars which are left in the centre have small velocities of approximately 10 \kms but
low accretion rates as they are essentially starved by their more massive siblings. Upon being
ejected from the halo the two most massive stars carry more than $10^5$ \msolar of gas out of the
halo dramatically altering the gravitational potential of the halo centre. \\
\indent The most massive stars exit the central region and their accretion rates decline
significantly (see \S \ref{seed}). It should be noted at this point that the stellar mass at the centre
dominates the gravitational potential relatively quickly after formation due to their very high accretion rates.
Hence, the stars are very unstable to ejections because there is little high
density gas left to provide a gravitational attraction. These ejected stars are hyper-velocity stars
with enough velocity to exit the halo (the escape velocity of the halo is only
20 \kms). These massive stars, final masses of 76123 \msolar and 44727 \msolar respectively,
will form a population of wandering massive black holes \citep[e.g.][]{Tremmel_2018}.\\
\indent We select a merger radius for all stars in our
simulations at 4 cell widths. Therefore, mergers and close encounters occur at the edge of our
resolution. The gravity solver in Enzo is a particle mesh gravity solver with an effective
  resolution of twice the cell spacing - therefore our gravitational resolution is two cell spacings.
Increasing our merger radius may eliminate these ejections by dulling
our ability to capture fragmentation. By setting our merging radius at this low value we
are marginally able to track these stars at the cost that these close encounters, and their
resulting kicks, are only marginally resolved. As we simulate only a single halo, albeit
at varying resolution and with different external backgrounds, our results can not be ascribed to
all haloes. Our results are qualitative in that ejections likely play a role when multiple massive
fragments form in close proximity but placing quantatative values on the fraction of haloes
which experience ejections or on the exact environmental conditions which either promote
or suppress ejections is outside the scope of this work.

\subsection{Feedback from the Pop III stars}
As the stars evolve their individual accretion rates determine whether they are able to sustain
a bloated envelope and remain a SMS or whether they contract to the main sequence and become
massive Pop III stars. In our subgrid implementation we set the critical threshold at 0.04 \msolarc/yr
and allow particles to remain as SMSs as long as any dips below this threshold are shorter that 1000
years \citep{Sakurai_2016}. Nonetheless, for our fiducial 100 J$_{21}$ runs none of the
particles that form are able to sustain accretion rates above 0.04 \msolarc/yr. Even in the Ref10
runs (where as we noted low resolution results in artificially high accretion rates)
the star transitions to a massive Pop III star rather than a SMS after approximately 300 kyr.
For the Ref14 and Ref18 cases the accretion
rates quickly fall below the critical threshold and the stars contract to the main sequence. As the
stars contract to the main sequence the radiation feedback switches from an effective temperature
of $\rm{T_{eff} = 5000 K}$ to $\rm{T_{eff} = 10^5 K}$. The resulting radiation spectrum switches from
predominantly infrared radiation to predominantly UV radiation.  In Figure \ref{MultiPlot} we plot the
temperature and electron fraction as a function of radial distance.
In all cases the profiles are centred on the most massive Pop III star which emits a spectrum
based on a blackbody temperature of $\rm{T_{eff} = 10^5\ K}$. The temperature of gas is initially
close to isothermal even up to the accretion radius of the Pop III star. The high gas densities
surrounding the stars are effective in regulating the central temperatures at close to $10^4$ K.
As the ionising radiation from the Pop III star starts to ionise and heat the surrounding gas the
temperature increases and has reached a value of around $6 \times 10^4$ K after 250 kyr.
The electron fraction (right hand panel of  Figure \ref{MultiPlot}) shows a sharp increase in the
electron fraction close to the star as the HII region expands and reaches a value of unity close to
1000 AU. However, the high gas column densities
successfully shield the gas and prevent the HII region expanding much beyond that, with the electron fraction
returning to values close to $10^{-5}$ at 3000 AU. 
In Figure \ref{TwoPanel} we plot slices through the electron fraction field showing the impact of the
ionising radiation from the last two snapshots shown in  Figure \ref{MultiPlot}. In both panels the
electron fraction reaches values well above 0.1 within the accretion radius of each Pop III star
with dense, neutral, gas sitting around the Pop III stars. The extent of the HII region in no more
than a few cells wide - a few thousand AU at most. The high density of the surrounding gas,
combined with the high infall rates limit the ability of the HII regions to expand. \\
\indent Upon contraction to the main sequence the Pop III star radii fall well below our resolution
limits. Given we set the accretion radius at 4 cells in radius we are therefore most likely
over estimating the accretion onto the Pop III star as the ability of the gas to discard angular
momentum at scales near the star can clearly not be captured by this work. The feedback scheme
implemented here scales with the mass of the star - we are therefore also over estimating the feedback
intensity from the stars. Even in over estimating the feedback it is nevertheless trapped close
to the star and hence our conclusion that feedback from Pop III stars in regions of high infall
does not negatively effect the infall is robust. \\
\indent We also note that the results found for the Pop III stars are consistent with those
  for present day massive star formation. Results from simulations of present day massive
  star formation \citep[e.g.][]{Peters_2010, Krumholz_2015b} show that fragmentation is less of an
  issue than was once believed. Similar to the results found here many authors (within the
  present day star formation community) have found that
  radiative feedback is successful at heating the gas thus suppressing fragmentation. Ionising
  radiation which can drive large HII regions has also been found to be
  unable to halt the accretion flow as long as the flow is maintained and the HII regions' expansion
  subdued \citep{Walmsley_1995, Peters_2011}.

\subsection{Towards forming a massive black hole seed} \label{seed}
Due to the high computational cost we are unable to evolve our highest resolution simulations, even
when running in optically thin mode, much beyond 250 kyr within a reasonable timeframe.
Nonetheless, for some of the lower resolution runs we can explore the future fate of these objects.
As can be concluded from Figures \ref{MassAccretionRate} and \ref{MassAccretionRate_Resolution}
after approximately 100 kyr the total mass of the \smartstars saturates and the accretion rate
also saturates. Furthermore, our examination of the stellar dynamics at the centre of the
potential when multiple interactions occur show that massive bodies can be flung out from the
centre of the halo thus severely diminishing their accretion rates.  This result is borne out in
Figure \ref{MassAccretionRate_All} where we plot the accretion rate up to 500 kyr after the
formation of the first \smartstar for all of the stars which survive to the end of the Ref18\_100J
and Ref18\_1000J\_OT simulations. Each of the
stars trace a unique accretion history. The two most massive stars in the Ref18\_1000J\_OT simulation
have high accretion rates until approximately 250 kyr after their formation. At this point, as already
discussed in \S \ref{ejections}, the stars get ejected from the central core and their accretion rates
drop dramatically. The Ref18\_100J simulation exhibits a similar event, a four body interaction at
140 kyr causes the ejection of the two surviving \smartstars from the system and with it a
significant drop in accretion as the \smartstars leave the high density region. Therefore,
while both the  Ref18\_100J and Ref18\_1000J\_OT simulations form massive Pop III stars, after
approximately 500 kyr the stars also get ejected due to multi-body interaction. These stars will
leave the halo, collapse into black holes but be unable to grow. \\
\indent In our highest resolution run, Ref20\_100J, there were fewer multibody interactions. In this
system a single massive Pop III star remains at the centre of the rapid inflow after 250 kyr. 
While the prohibitive computational cost of this
high resolution system leaves us unable to evolve this system for more than 250 kyr, at this
evolutionary stage this single surviving Pop III star is likely to remain at the centre of the
potential accreting at relativity high accretion rates ($\dot{M} \sim 10^{-2}$ \msolarc/yr) and
ending its life with a mass of approximately $2 \times 10^4$ \msolarc.
This Pop III star will then directly collapse into a black hole of the same mass 
continuing to accrete mass. If the accretion rate at this time remains at levels close
to $\dot{M} \sim 10^{-2}$ \msolarc/yr this value will exceed the canonical Eddington rate 
by a factor of close to 100. Feedback from the black hole will likely quickly regulate the
accretion rate. Nonetheless, the black hole will be centred at the apex of the gravitational
potential supported by a high inflow rate and therefore likely to experience high or super
critical accretion rates. Therefore, for the case of the 100 J$_{21}$ at the highest resolution
we find that conditions, for this halo, are optimal for forming a rapidly accreting black hole
seed.\\

\section{Summary \& Discussion}  \label{Sec:Discussion}

In this study we have tracked the formation of SMSs and massive Pop III stars in haloes experiencing
high inflow of up to 1 \msolarc/yr. We have used two different LW backgrounds of 100 J$_{21}$ and
1000 J$_{21}$ to create the necessary environmental conditions. We find that with a background of
100 J$_{21}$ a massive Pop III star forms. The most massive star to form under such a background
initially experiences sufficient accretion rates to allow its envelope to expand and it forms a
SMS, however, the star cannot maintain the required accretion rates and after less than 100 kyr
contracts to the main sequence. The 1000 J$_{21}$ is able to induce a critical accretion rate for
at least 250 kyr and a SMS forms. Under both backgrounds mild fragmentation is observed within the
collapsing gas. The birth of multiple fragments within the tight confines leads to tight binaries
and subsequent coalescences but also to violent, multibody, dynamical interactions leading to the
ejections of stars in many cases. In our highest resolution simulation with a background of 1000
J$_{21}$ a series of close encounters results in the ejection of the two SMSs from the centre of the
halo. The ejection results in hugely diminished accretion rates and the stars contract to the main
sequence on their way out of the halo. In the highest resolution simulation with a background of
100 J$_{21}$ multiple interactions between multiple stars are fortuitously avoided (due to mergers) and the
systems settles with one massive Pop III star accreting at the centre of the halo. \\
\indent The very high inflow rates suppress the impact of the ionising radiation from the massive
Pop III stars which form. HII regions do form around the Pop III stars but surrounding high density
gas effectively absorbs the emitted radiation arresting the propagation of the HII region much beyond a
few cell widths from the Pop III stars. Our maximum resolution is of the order of 100 AU and so while
this is marginally sufficient to capture SMS formation and the accretion onto the outer envelope
of a SMS it is insufficient to follow the accretion onto the surface of a Pop III as it contracts
on the Kelvin-Helmholtz timescale. In our scheme we assume that gas that is flowing radially inward
at close to our maximum resolution makes it onto the surface of the star. There are clearly inherent
limitations to this approach and we are likely over estimating the accretion rate onto Pop III stars.
Nonetheless, we observe that feedback from these, very massive, Pop III stars is unable to evacuate
a significant HII region due to the high mass inflow and we conclude that this is a robust result
from our simulations. A further caveat of our simulation setup is the imposition of a LW background.
The background is necessary to suppress \molH formation and allow for the formation of atomic
cooling haloes with sufficient inflow rates to investigate the formation of a SMS. However, the
environment is somewhat artificial as the nearby star forming galaxies which would in reality
create the LW radiation are absent along with their gravitational impact on the simulated galaxy. These
nearby galaxies may induce torques which could potentially suppress or enhance the accretion rates
and also facilitate future mergers. We will investigate more, cosmologically realistic, environments
in an upcoming study. \\
\indent \cite{Chon_2018} undertook a recent study where they explored the formation of SMSs drawn from
realistic cosmological simulations, albeit also with the imposition of an artificial LW background to
suppress \molHc. \cite{Chon_2018} investigated two collapsing haloes - one which was subject to
significant dynamical torques (filamentary collapse) and one where the tidal field was relatively weak
(spherical collapse). In both cases they follow the evolution of the SMSs for 100 kyr at very
high resolution. Similar, to our findings they report the formation of SMSs with accretion rates
exceeding the critical rate. They find that mild fragmentation is also prevalent in their simulations
with more than 10 stars forming in their filamentary galaxy. In their simulations the primary
protostar maintains a super-critical accretion rate for the majority of the 100 kyr. They do not
see any ejections
of stars from their galaxy in the first 100 kyr. However, this timescale is relatively short -
representing only approximately 10\% of the lifetime of the most massive primary star. Ejections in
our simulations only became evident after approximately 200 kyr as the interactions of the stars
take time to develop with the system undergoing significant evolution in the first 200 kyr with new
stars forming and mergers common. Nonetheless, the two sets of simulations are broadly consistent
with protostars showing super-critical accretion rates and the formation of multiple stars in
both cases. \\
\indent Our simulations show that the formation of both massive Pop III stars and SMSs is
viable in regions
of high accretion. Feedback in both cases has little or no negative impact on the accretion flow
at the scales probed in our work. Addressing the future fate of these objects is less trivial.
As discussed above the SMSs which form and evolve with rates above the critical limit in the
1000 J$_{21}$ simulations both experience violent dynamical interactions which result in their
expulsion from the halo. While these seeds will likely go on to form massive seed black holes
they will be far from the halo centre and likely unable to accrete any further gas. In the 100
J$_{21}$ case a single Pop III star forms at the centre of the collapse and while its accretion rate
is sub-critical for SMS expansion the very high accretion rates would be significantly
super-Eddington for a black hole of comparable mass. If this massive Pop III star directly collapses
to a black hole and remains at the centre of a strong accretion flow the conditions are very
favourable for achieving super critical accretion if the angular momentum of the gas can be
shed efficiently \citep{Begelman_2017, Sugimura_2018}. It could also be that magnetic fields
induced by the initially magnetised gas and amplified by the small scale dynamo effect
\citep{Schleicher_2010b, Sur_2010, Turk_2012, Schober_2012, Schober_2013} induce angular momentum
transport helping to maintain very high accretion rates. \\
\indent Finally, the lighter levels of fragmentation in the 100
J$_{21}$ are a further attraction  as this leads to a lower probability of ejection from the halo
something that may effect larger systems which may suffer higher levels of fragmentation. A more
thorough exploration of the formation of massive seeds will require a more realistic treatment of the
dynamical environment in which the conditions for massive star formation becomes possible.

\section*{Acknowledgements}

J.A.R. acknowledges the support of the EU Commission through the
Marie Sk\l{}odowska-Curie Grant - ``SMARTSTARS" - grant number 699941.
Computations described in this work were performed using the 
publicly-available \enzo code (http://enzo-project.org), which is the product of a collaborative 
effort of many independent scientists from numerous institutions around the world.  Their 
commitment to open science has helped make this work possible. The freely available astrophysical 
analysis code YT \citep{YT} was used to construct numerous plots within this paper. The authors 
would like to extend their gratitude to Matt Turk et al. for an excellent software package.
J.A.R. would like to thank Lydia Heck and all of the support staff involved with Durham's COSMA4
and DiRAC's COSMA5 systems for their technical support. This work was supported by the Science
and Technology Facilities Council (grant numbers ST/L00075X/1 and RF040365). This work used the
DiRAC Data Centric system at Durham University,  operated  by  the  Institute  for  Computational
Cosmology on behalf of the STFC DiRAC HPC Facility  (www.dirac.ac.uk). This equipment was funded
by BIS National E-infrastructure capital grant ST/K00042X/1, STFC capital grant ST/H008519/1,
and STFC DiRAC Operations grant ST/K003267/1 and Durham University.  DiRAC is part of the
National E-Infrastructure. The authors also wish to acknowledge the SFI/HEA Irish Centre for
High-End Computing (ICHEC) for the provision of computational facilities and support.
Furthermore, the authors would like to thank John Wise and Marta Volonteri for useful discussions and
comments on earlier manuscript drafts. Finally, the authors would like to thank the referee
for a constructive report.  
\noindent

\bibliographystyle{mn2e_w}
%\bibliography{../BIBTEX/mybib}

\bsp	% typesetting comment

\label{lastpage}
\end{document}